\documentclass[reprint,amsmath,amssymb,aps,showkeys]{revtex4-2}

\usepackage{graphicx}
\usepackage{bm}      
\usepackage{amssymb,amsmath,amsthm}
\usepackage{braket}
\usepackage{dsfont}
\usepackage{color}

\newcommand{\abs}[1]{\left\vert#1\right\vert}
\renewcommand{\atop}[2]{\genfrac{}{}{0pt}{}{#1}{#2}}

\renewcommand{\i}{\mathrm{i}}

\newcommand{\1}{\mathds{1}}

\renewcommand{\Re}[1]{\mathbb{R}\mathrm{e}\left\{#1\right\}}
\renewcommand{\Im}[1]{\mathbb{I}\mathrm{m}\left\{#1\right\}}

\renewcommand{\H}{\mathcal{H}}

\begin{document}

\title[Exciton pairs.]{Exciton pairs coupled via the long-living phonons and their superfluorescent markers.}

\author{Vladimir Al. Osipov}
\affiliation{Institute for Advanced Study in Mathematics, Harbin Institute of Technology, 92 West Da Zhi Street, Harbin, 150001, China}
\affiliation{Suzhou Research Institute, Harbin Institute of Technology, 500 South Guandu Road, Suzhou, 215104, China\\ Email: Vladimir.Al.Osipov@gmail.com}

\author{Boris Fainberg}
\affiliation{Faculty of Science, H.I.T.-Holon Institute of Technology, 52 Golomb Street, POB 305, Holon, 5810201, Israel}

\begin{abstract}  A system of several Wannier-Mott excitons interacting with phonons in a bulk material is considered. We show that strong exciton-phonon coupling causes the formation of a coherent two-exciton state -- the exciton pair. Unlike the biexcitons, where the photons play the role of force carrier, the exciton pair is formed via entanglement with the long-living phonon mode: (i) The essentially multi-particle theory requires excitons (cobosons composed of an electron and a hole) to satisfy the mixed Bose-Fermi statistics; (ii) This allows us to formulate a system of non-linear dynamic equations, using the multiconfiguration Hartree method applied to the Fr\"ohlich Hamiltonian. The system of equations possesses a stationary solution, which, for the case of a single exciton, describes the excitonic polaron and corresponds to the exciton pair in the two-exciton case. We also compare the fluorescent spectra of exciton polarons and exciton pairs estimated from our theory with those observed in experiments on room-temperature superfluorescence (collective emission of fluorescent light) in hybrid perovskites to give an additional insight into the superfluorescence phenomenon.
\date{\today}
\end{abstract}


\maketitle

Semiconductors in all dimensions, ranging from quantum dots to bulk crystals, support bound electrically neutral electron-hole pair quasiparticles termed as Wannier-Mott excitons. They are formed due to the attractive Coulomb interaction of the charges and have an internal structure of a hydrogen atom. Wannier-Mott excitons (below simply excitons) possess a large Bohr radius compared to the crystal lattice constant, a high diffusion constant and a low exciton binding energy, yet stable at room temperature~\cite{Pyshkin_2018, ExcitonReview2021, Subhabrata2025}. 

Spatial fluctuations of the electric charge density of a charged particle can polarise the surrounding crystal lattice, i.e. interact with the lattice vibrations. The well-studied example of this type is polaron~\cite{Emin_2012}, the fermionic quasiparticle composed of an electron ``dressed'' with the carrier-induced lattice polarisation. While the polaron theory was originally developed for electrons, there is much experimental and theoretical evidence that the neutral bosonic excitons form excitonic polarons (EP)~\cite{Schmidt69, IB1983, SinghMatsui1987, ZhChJF2024}. In this letter, we show that the interaction of excitons with long-living phonons can lead to the existence of a {\it two-excitonic stable coherent state}, which we call exciton pairs or exciton molecules (EM). This state should not be mixed with biexcitons, which are composed of two holes and two electrons held together due to Coulomb forces~\cite{Singh2Dbiexciton1996, LiMaWangFangLuoLi+2020+2001+2006}. 

The system of Wannier excitons interacting with phonons is typically considered within the effective mass approximation, when the exciton motion as a whole is separated from its internal structure. In this approximation, the exciton energy is a sum,  $W_q=E_{int}+q^2/2 M_{ex}$, of the internal exciton energy $E_{int}$ and quadratic in the momentum $\bm q$ kinetic term~\cite{E1957, Singh1994}, where $M_{ex}$ is an effective exciton mass, $M_{ex}=m^*_h+m_e^*$. In a polar crystal, the longitudinal optical (LO) phonons have the strongest cohesion to the exciton. Excitons interacting with the LO phonons are described by the Fr\"ohlich Hamiltonian~\cite{Frohlich01071954} ($\hbar$ is set to 1 along the text):
\begin{multline}
\hat{\H}= \omega_{ph}\textstyle\sum_{\bm q}\hat{b}_{\bm q}^\dag \hat{b}_{\bm q}+\textstyle\sum_{\bm q}  W_q\hat{B}_{\bm q}^\dag \hat{B}_{\bm q}
\\+  \textstyle\sum_{\bm k,\bm q}D(k) \hat{B}_{\bm q}^\dag \hat{B}_{\bm q-\bm k}(\hat{b}_{-\bm k}^\dag + \hat{b}_{\bm k})+\hat{\H}_{bath}+\hat{\H}'.
\label{eq.:H_03}
\end{multline}
The operators $\hat{b}_{\bm q}^\dag$, $\hat{b}_{\bm q}$ are creation/annihilation bosonic operators of a phonon in the mode $\bm q$. In a good approximation, the LO phonon frequency is a constant $\omega_{ph}$. The third term in eq.~(\ref{eq.:H_03}) is the Fr\"ohlich type of exciton-phonon interaction~\cite{AL2022, Singh1994}, which describes the completely inelastic scattering of a phonon on an exciton with the Fr\"ohlich amplitude $D(k)$. The term $\hat{\H}_{bath}$ includes the interaction of the phonon modes with the thermal bath and $\hat{\H}'$ describes the exciton-light coupling (they will be specified later).

The excitons are compound particles made of two fermions (cobosons~\cite{CBD2007}) and thus the operators of exciton creation/annihilation ($\hat{B}^\dag_{\bm q}$ and $\hat{B}_{\bm q}$) with different wavevectors $\bm q$ and $\bm q'$ satisfy the Bosonic commutator relations. The correction to the Bose commutator calculated at $\bm q'=\bm q$ is proportional to the concentration of excitations and usually neglected at low concentrations. Proper calculation of the anticommutator $[\hat{B}_{\bm q},\hat{B}^\dag_{\bm q}]_+$, when the operators are expressed in terms of electron and hole operators, shows that the correction to the Fermi statistics is proportional to the overlap of the exciton wave packets. Unless we are concerned with the exciton Bose condensates (we consider high temperatures) or excitons generated in a small volume (we consider bulk material), these corrections are small for all reasonable exciton concentrations, and we stick to the mixed Bose-Fermi statistics of the exciton operators,  
\begin{equation}\label{paulions}
[\hat{B}_{\bm q},\hat{B}^\dag_{\bm q}]_+=1,\mbox{ but }[\hat{B}_{\bm q'},\hat{B}^\dag_{\bm q}]=0,\mbox{ for } \bm q'\ne \bm q.
\end{equation}

{\it Equations of motion.}
To explore the dynamics of the exciton-phonon system we make use of the multiconfiguration Hartree approach~\cite{Davydov1973, MEYER199073, Miller2002, AR2017, Zhao2022, Osipov_Fainberg23PRB}, which is known to be useful for describing the dynamics of quantum systems containing fields of different nature. At the derivation of the equations of motion, the vibrational part of the wavefunction is expanded over the time-dependent basis of coherent states (Davydov Anzats~\cite{Davydov1973}), parametrised by the complex-valued eigenvalue $\sigma_{\bm k}$ of the corresponding phonon operator: $\hat{b}_{\bm k}\ket{\sigma_{\bm k}}=\sigma_{\bm k}\ket{\sigma_{\bm k}}$. Neglecting the higher-order powers of $\hat{B}_{\bm q}^\dag$, the multi-exciton wave-function is chosen in the form (bold $\bm \sigma$ stands for the set of all $\sigma_{\bm k}$ and $\ket{0_{\bm q}}$ is the zero $\bm q$-exciton state)
\begin{equation}\label{exstate0}
\ket{\Psi(t)}=  \textstyle\prod_{\bm q}\left(G_{\bm q}(t)+ C_{\bm q}(t)  \hat{B}_{\bm q}^\dag\right)\ket{0_{\bm q}} \ket{\bm \sigma(t)},
\end{equation}
subject to the normalizing condition, $\abs{G_{\bm q}}^2+ \abs{C_{\bm q}}^2 =1$.

Variation of the Schr\"odinger equation over $C$, $G$ and $\sigma$ generates a system of equations of motion~\cite{AR2017, Miller2002, MEYER199073}: the equation of a dumped classical oscillator driven by the excitonic field for $\sigma_{\bm k}$; and the non-linear (quintic) extended Schr\"odinger equation~\cite{Aklan_2021} for the expansion coefficients $G_{\bm q}$ and $C_{\bm q}$ with the non-linear terms modulated by the vibration strength amplitude $\alpha(k)=D(k)(\sigma^*_{-\bm k}+\sigma_{\bm k})$. It is convenient to write down the set of equations in terms of the complex-valued polarization $Q_{\bm q}\equiv \bra{\Psi }\hat{B}_{\bm q}\ket{\Psi }=G^*_{\bm q} C_{\bm q}$ and the relative population $J_{\bm q}=\bra{\Psi )}\hat{J}_{\bm q}\ket{\Psi }\equiv \bra{\Psi )}[\hat{B}^\dag_{\bm q},\hat{B}_{\bm q}]/2\ket{\Psi }=(\abs{C_{\bm q}}^2-\abs{G_{\bm q}}^2)/2 $. Then, the equations of motion read 
\begin{eqnarray}
\dot\sigma_{\bm k}&=& -\i  \omega_{ph}  \sigma_{\bm k} -\gamma\sigma_{\bm k} -\i D(k)\textstyle\sum_{\bm q}  Q^*_{\bm q}Q_{\bm q+\bm k};\label{eqsigma}\\
\dot Q_{\bm q}&=&-\i  W_q  Q_{\bm q}  +2\i J_{\bm q}\textstyle\sum_{\bm k}\alpha(\bm k)  Q_{\bm q-\bm k};\label{eqQ}\\
\dot J_{\bm q}&=& 2 \Im{ Q^*_{\bm q}\textstyle\sum_{ \bm k} \alpha(\bm k) Q_{\bm q-\bm k}},\label{eqJz}
\end{eqnarray}
The dumping term $-\gamma\sigma_{\bm k}$ in eq.~(\ref{eqsigma}) accounts for energy dissipation into the thermal bath (the term $\hat{\H}_{bath}$ in eq.~\ref{eq.:H_03}). By direct checking, one can show that the normalisation condition and the conservation of the total number of excitations $\sum_{\bm q} J_{\bm q}$ are satisfied. The energy functional can be read off from the Hamiltonian (eq.~\ref{eq.:H_03}),
\begin{equation}\label{EnergyConserva}
E  =\textstyle\sum_{\bm q}\left(\omega_{ph}\abs{\sigma_{\bm q}}^2+ W_q J_{\bm q}\! +\! 2\Re {\textstyle\sum_{\bm k}\alpha(\bm k)Q_{\bm q} Q^*_{\bm q+\bm k}}\right).
\end{equation} 
Presence of dissipative terms in eq.~(\ref{eqsigma}) violates conservation of energy. However, {\it for the stationary solution}, when $\sigma_{\bm k}^{(st)}= -\frac{D(k)}{\omega_{ph}-\i \gamma} \sum_{\bm q}  Q^*_{\bm q}Q_{\bm q+\bm k}$, {\it the energy reaches the minimal value and its time derivative nullifies}. Emphasise that the steady state can be reached when, at least, a few terms in the sum $\sum_{\bm q}  Q^*_{\bm q}Q_{\bm q+\bm k}$ ($\bm k\ne 0$) are in phase. In other words, the exciton wave packet or the wave packets of several excitons have to be completely or partially entangled with the long-living phonon mode (LLPM). Two special stationary solutions corresponding to EP and EM are discussed in the next section.
  
\begin{figure}
\begin{center}
\begin{tabular}{lc}
a)&\includegraphics[scale=0.38]{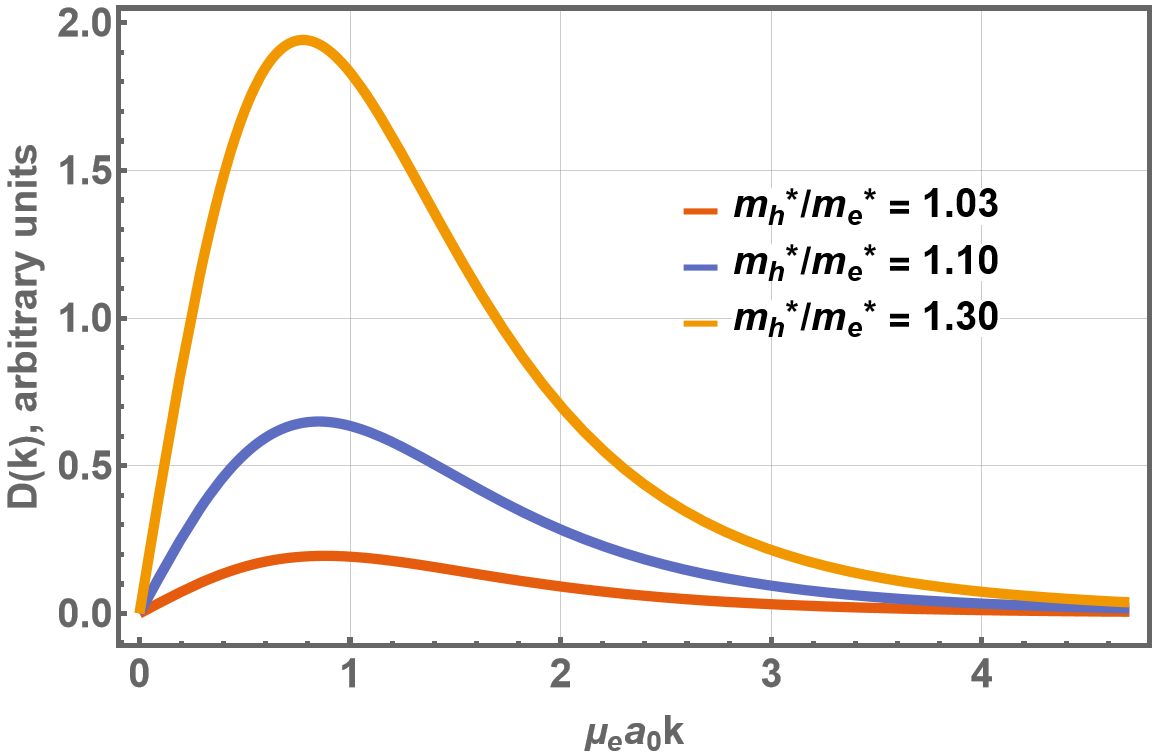}
\end{tabular}
\end{center}
\caption{\small \label{Fig1}The Fr\"ohlich exciton-phonon interaction amplitude in thin films plotted vs. the dimensionless wavevector $\mu_e a_0k$, where $a_0$ is the exciton Bohr radius, and $\mu_e$ is the ratio of the electron effective mass to the exciton reduced mass, $\mu_e=m_h^*/(m_e^*+m_h^*)$. The position of maximum is $\mu_e a_0k_0\sim 0.8$ and almost independent on the ratio $m_h^*/m_e^*$.
}
\end{figure}

\begin{figure}
\begin{center}
\begin{tabular}{lc}
a)&\includegraphics[scale=0.70]{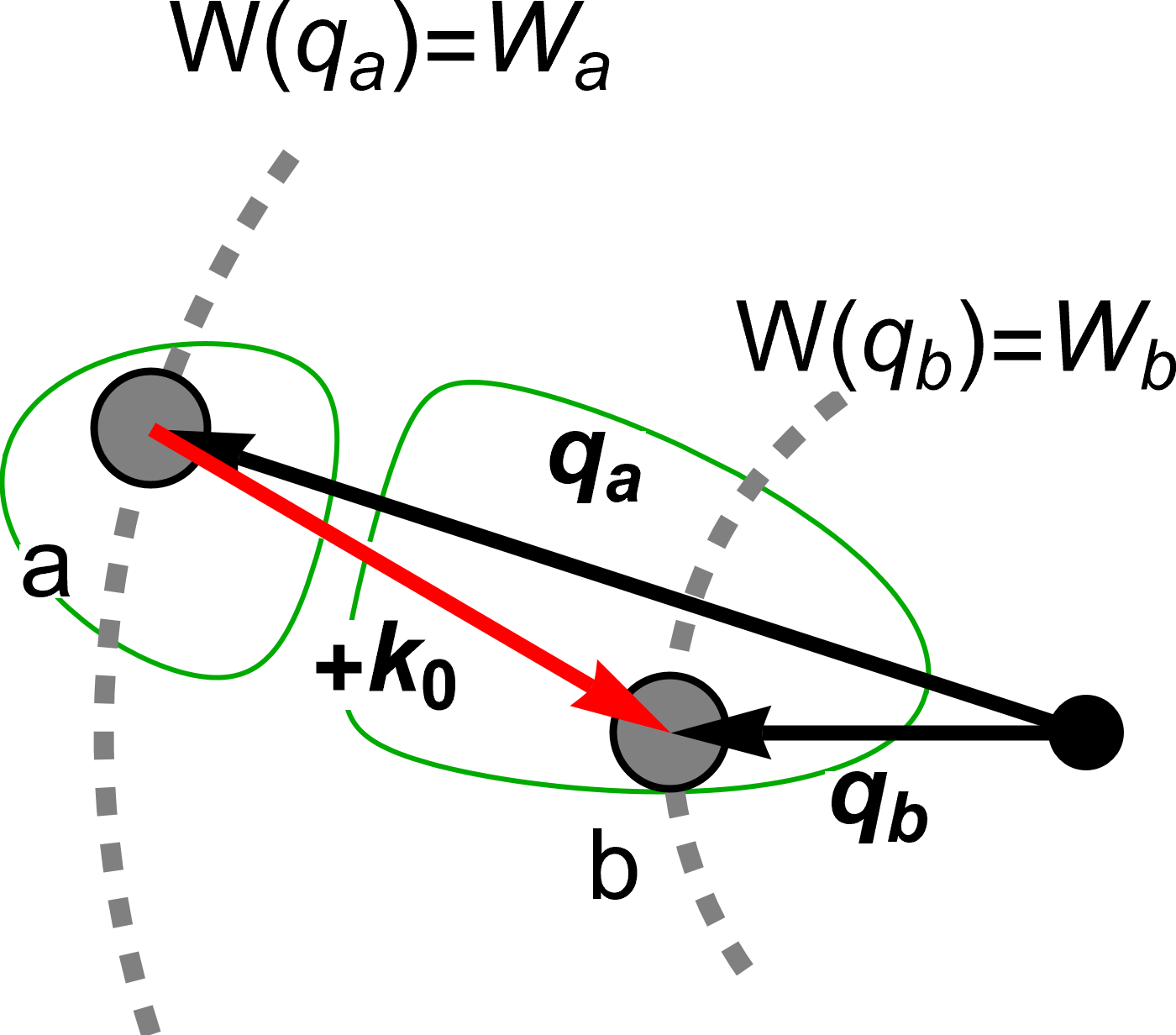}\\
b)&\includegraphics[scale=0.33]{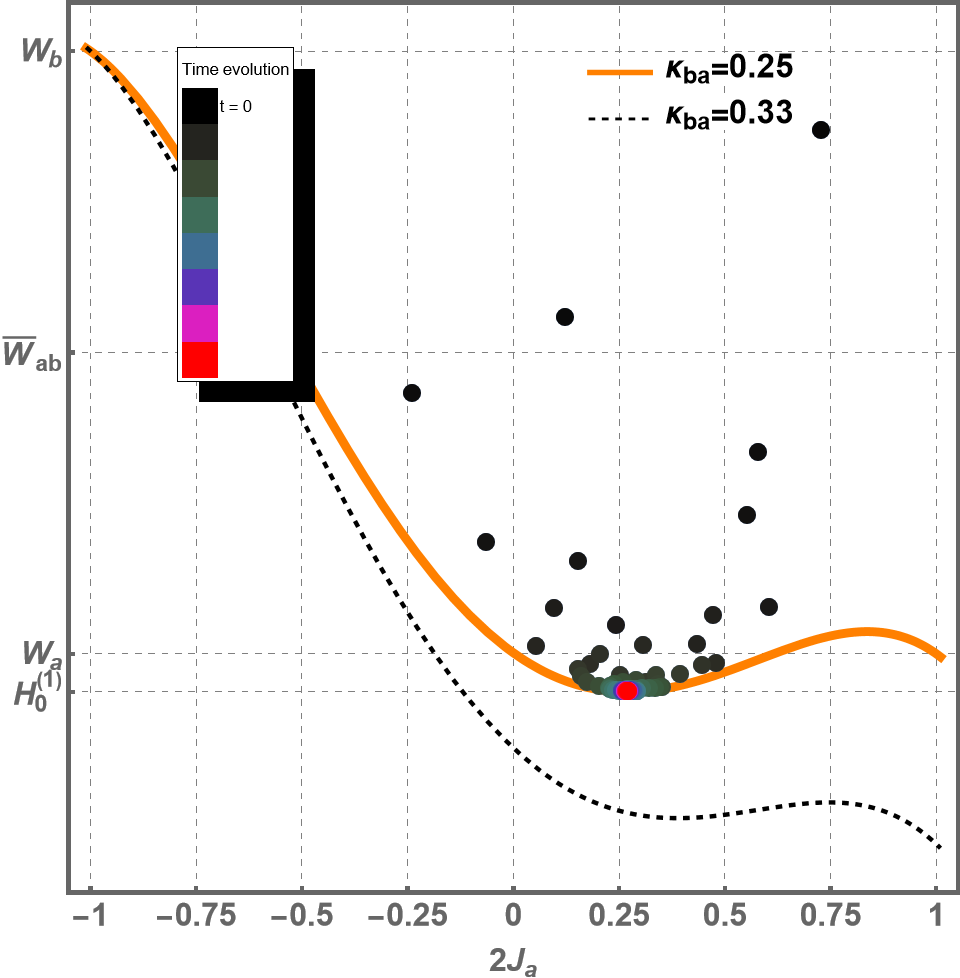}
\end{tabular}
\end{center}
\caption{\small \label{Fig2}(a) In the truncated model the exciton wave packet is effectively represented by two quantum states a and b (with energies $W_a$ and $W_b$ and wavevectors $\bm q_a$ and $\bm q_b$, respectively) connected by the  vector $\bm k_0$ of LLPM. (b) The energy $H^{(1)}$ (orange curve, eq.~\ref{energyWa}) and numeric solution (coloured dots) of the equations of motion (eqs.~\ref{eqsigma},~\ref{eqJz},~\ref{eqQ} for the case depicted in Fig. a) at $\kappa=0.25$. The dots show the energy (eq.~\ref{EnergyConserva}) and the relative population $J_a$ of the exciton-phonon subsystem during its evolution. Colours depict the time step: from black at small times to red at larger times. The system evolves from some non-stationary high-energy state towards the stationary solution corresponding to the energy minimum. The function $H^{(1)}(J_a)$, plotted at $\kappa_{ba}=0.33$ (dashed black), is given for comparison. 
}
\end{figure}

{\it The stable stationary solutions for one and two excitons.}
We perform an analytical analysis of the stationary solutions for the truncated version of the equations of motion. The Fr\"ohlich interaction amplitude $D(k)$ for neutral excitons~\cite{Toyozawa59,Fainberg_Osipov24JCP} nullifies at $k=0$ and reaches maximum at some wavevector $k_0$. Its value can be estimated from its analytic expression\footnote{
\begin{multline*} 
D(k)=\omega_{ph}\sqrt{\frac{\alpha_{LO}}{V\sqrt{2m^*_e \omega_{ph}}}}\;\frac{4\pi}
{k}\\
\left[
 \left(1+\left(\frac{\mu_e a_0 k}{2}\right)^2\right)^{-\xi} 
-\left(1+\left(\frac{m^*_h  \mu_e a_0 k}{2m^*_e}\right)^2\right)^{-\xi}
\right],
\end{multline*} 
where $a_0$ is the exciton Bohr radius, $\mu_e$ is the ratio of the electron effective mass and the exciton reduced mass, $\mu_e=m_h^*/M_{ex}$, $\alpha_{LO}$ is the coupling constant, and $V$ is  the crystal volume. The exponent $\xi=2$ for the bulk material~\cite{Toyozawa59} and changes to $\xi=3/2$ for the geometry of thin films~\cite{Fainberg_Osipov24JCP}.} to give with a good accuracy $k_0= \mathit{const.}/a_0$, where $a_0$ is the exciton Bohr radius and the constant is almost independent on $m_h^*/m_e^*$ ratio, and is slightly influenced by the material shape: for bulk $\mathit{const.}=3.8$ and reaches $4$ for the thin film geometry (see Fig.~\ref{Fig1}). Moreover, considering that the non-linear terms in the equations of motion become important for significant amplitudes only, we simplify our theoretical description and replace the whole exciton packet in the $\bm q$-space with only two quantum states connected by the vector $\bm k_0$ (states a and b in Fig.~\ref{Fig2}a), and assume that most of the quantum density is distributed over these two quantum states.

For two excitons and a single LLPM (Fig.~\ref{Fig2}a) the truncated version of the equations of motion read 
\begin{eqnarray}
\dot\sigma_\pm &=& -\i  \omega_{ph}  \sigma_\pm -\gamma\sigma_\pm -\i D  \left\{\!\atop{Q^*_a Q_b + Q^*_d Q_c}{Q^*_b Q_a + Q^*_c Q_d}\!\right\}; \label{eqsigmat}\\
\dot Q_a&=&-\i  W_a  Q_a  +2 \i J_a \alpha^*Q_b;\label{eqQt} \\
\dot J_a&=&  -2\Im{\alpha Q_a Q_b^*},
\label{eqJzt}
\end{eqnarray}
where $\sigma_\pm\equiv \sigma_{\pm\bm k_0}$, $D\equiv D(\bm k_0)$,  and $\alpha=D(\sigma^*_{-} + \sigma_{+})$, and we use the convention that index $i=a,b,c,d$ replaces $\bm q_i$. Equations for the state $b$ are obtained by the replacements $a\leftrightarrow b$ and $\alpha^*\to\alpha$  in eqs.~(\ref{eqsigmat} --~\ref{eqQt}). For $d$ and $c$ states the replacement is  $a\to c$, and $b\to d$. 

To consider the case of a single exciton (Fig.~\ref{Fig2}a) we set $Q_c=Q_d=0$. The long-time solution for the phonon amplitudes $\sigma_\pm$ can reach the steady state value $\sigma_\pm^{(st.)}=-\frac{D}{\omega_{ph}-\i \gamma}Q_a^* Q_b$ (eq.~\ref{eqsigmat}) under the phase-locking condition, i.e. when the product $Q_a^*Q_b$ is constant or, at least, slowly changes in time. The steady state vibration strength $\alpha^{(st)}= - 2\omega_0 Q_a^*Q_b$, with the energy scale $\omega_0\equiv \frac{\omega_{ph}D^2}{\omega_{ph}^2+\gamma^2}$, typical in the EP theory. For the stationary regime ($\sigma_\pm\to\sigma_\pm^{(st.)}$) under the conservation of the number of excitations condition ($J_a=-J_b$) the energy functional (eq.~\ref{EnergyConserva}) becomes a function of a single variable~$J_a$:
\begin{equation}\label{energyWa}
E\to H^{(1;a,b)}[J_a] = \overline{W}_{ab} - \omega_0\kappa_{ba}  J_a - \frac{\omega_0}{8}(1-4J_a^2)^2,
\end{equation}
where $\overline{W}_{ab}\equiv \frac{W_a+W_b}{2}$ is the mean energy and $\kappa_{ba} \equiv \frac{W_b-W_a}{\omega_0}$ is the  parameter describing deviation from the mean value. For eqs.~(\ref{eqsigmat}--\ref{eqQt}) the functional (eq.~\ref{energyWa}) plays a role of Lyapunov function, i.e. its local minimum is reached on their stable solution. The function $H^{(1)}$ has three special points. Two of them correspond to free excitons:  $H^{(1)}\to  W_{a/b}$ at $J_a=\mp1/2$. The local minimum is realised at $ J_{b,0}=-J_{a,0}=\kappa_{ba}(1+\kappa_{ba}^2)/2+\mathcal{O}(\kappa_{ba}^5)$ with the energy $H_0^{(1)} = \overline{W}_{ab} - \frac{\omega_0}{8}(1+\kappa_{ba}^2)^2+\mathcal{O}(\kappa_{ba}^5)$. The minimum exists in a narrow band $\abs{\kappa_{ba}}< 2/\sqrt{27}\sim0.38$ (Fig.~\ref{Fig2}b). Therefore, at a fixed $\kappa_{ba}$ the EP particle is stable and coexists with the free excitons. The effective potential barrier between this two exciton states is $\frac{ \omega_0}{8}\left(1-4\abs{\kappa_{ba}}+\mathcal{O}(\kappa_{ba}^2)\right)$ and becomes negligible at the edges $\abs{\kappa_{ba}}\simeq0.38$. The EP solution exists for geometries with almost arbitrary mutual orientations of the vectors $\bm k_0$, and $\bm q_a$ (Fig.~\ref{Fig2}a). The non-collinearity can cause angular slip of the wave packet in the momentum space until the configuration $\bm k_0\parallel\bm q_a$ is reached. The latter configuration corresponds to the lowest value of the EP energy $H_0^{(1)}$.

\begin{figure}
\begin{center}
\begin{tabular}{lc}
a)&\includegraphics[scale=0.55]{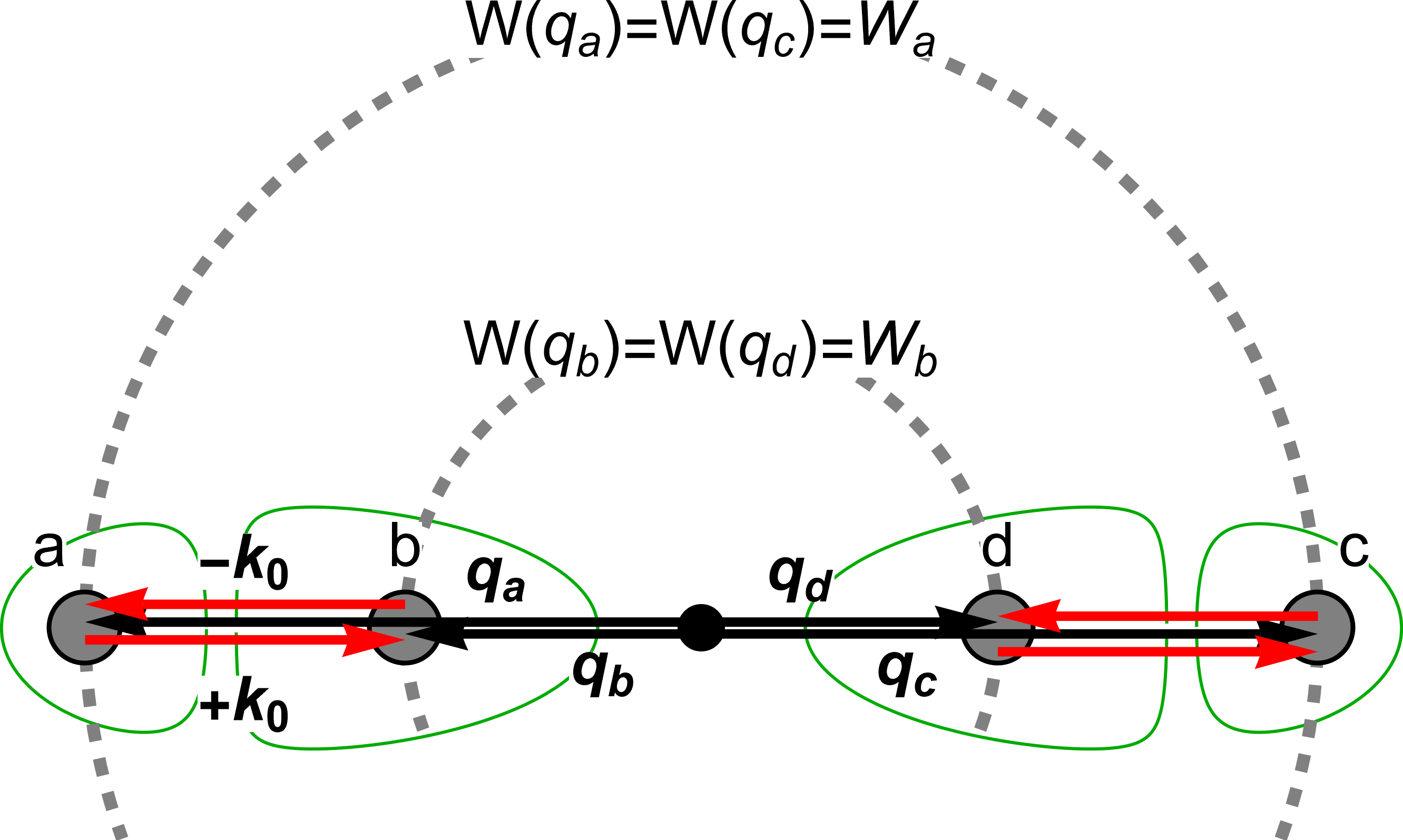}\\
b)&\includegraphics[scale=0.36]{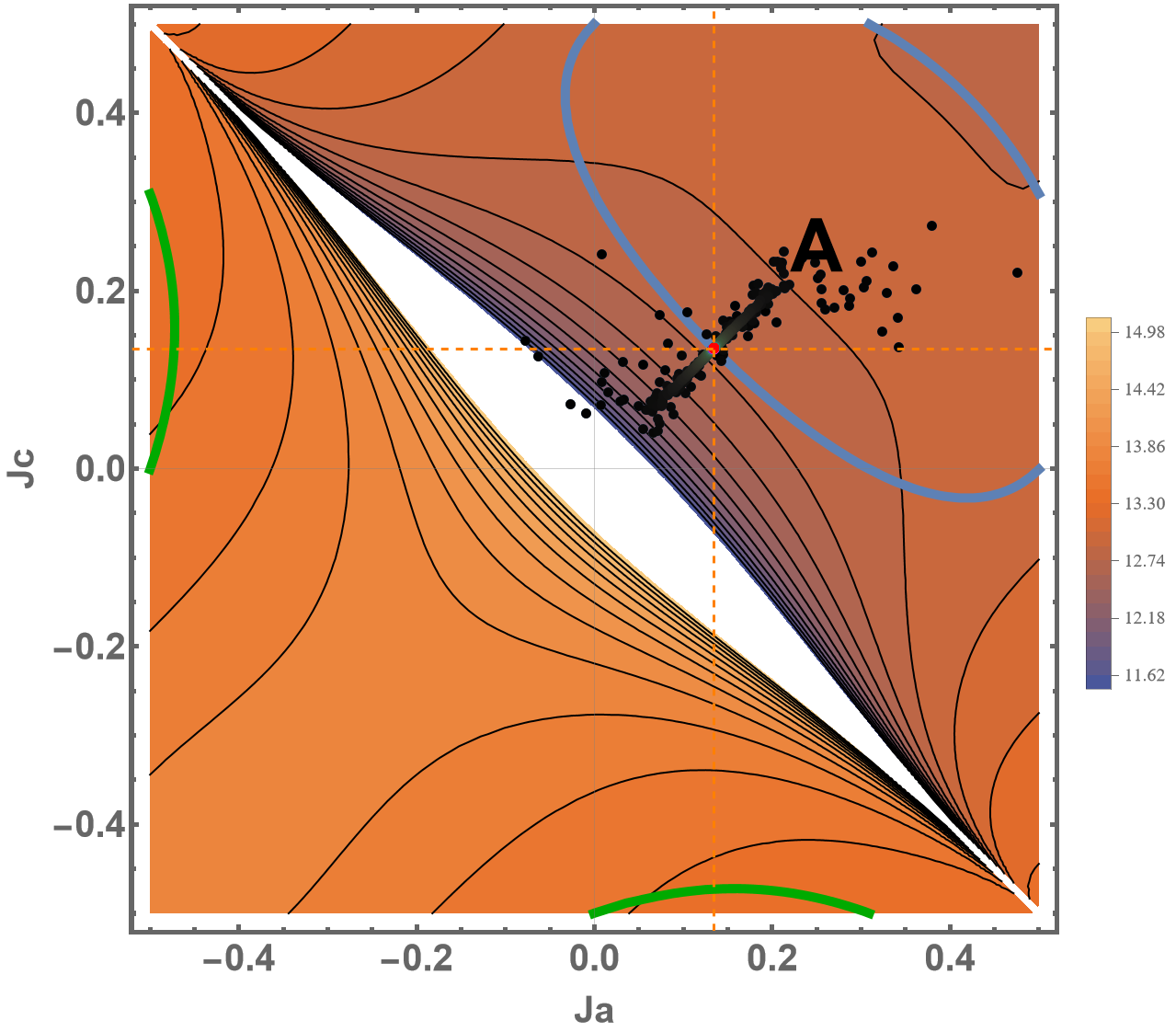}
\end{tabular}
\end{center}
\caption{\small \label{Fig3} (a) The MSEM is composed of two excitons entangled with one and the same LLPM. (b) The energy of EM $H^{(2)}$ (eq.~\ref{Esemicl4p1}) plotted as a function of the occupations $J_a$, and $J_c$ at $\kappa_{ba}=\kappa_{dc}=0.49$ (see the scheme in Fig. a, but with unequal energies $W_a\ne W_c$, and $W_b\ne W_d$) and numeric solution of the equations of motion  (the dots' colours were chosen as in Fig.~\ref{Fig2}b). In the region ``A'' surrounded by blue solid curve $0<\cos\Delta\le 1$, while in the regions surrounded by green curves $-1\le\cos\Delta<0$. The stationary state solution (crosspoint of thin dashed orange lines showing the stationary values $J_{c,1}=\kappa_{dc}/4$, and $J_{a,1}=\kappa_{ba}/4$) corresponds to the energy minimum at the blue curve, where $\cos\Delta=1$ . 
}
\end{figure}

To address the regime of high concentrations of excitations we turn our attention to the multiparticle model: two exitons with yet arbitrary energies are entangled with one and the same LLPM. Accounting of the quantum states c and d generates an additional term in the expression for the stationary vibration strength: $\alpha^{(st.)}=-2 \omega_0 \left(Q_a^*Q_b+ Q_d^* Q_c\right)$. The conservation of excitations, reflected in the relations $J_a=-J_b$ and $J_c=-J_d$, allows to write the energy as a function of $J_a$, $J_c$ and the phase $\Delta=\phi_a-\phi_b+\phi_c-\phi_d$ (where $\phi_i$ is the time-dependent phase of $Q_i$):
\begin{multline}\label{Esemicl4p1}
E\to H^{(2)}[J_a,J_c]= H^{(1;a,b)}[J_a]+H^{(1;c,d)}[J_c]\\- \frac{ \omega_0}{4} (1-4J_a^2)(1-4J_c^2) \cos\left[\Delta(J_a,J_c)\right].
\end{multline}
For the unlocked phase $\Delta$, the last term in eq.~(\ref{Esemicl4p1}) oscillates fast. Due to self-averaging, the energy effectively splits into two EP contributions, which must be minimized independently. In the phase-locked regime, i.e. when $\Delta$ reaches the stationary value $\Delta=0$, the energy $H^{(2)}$ has to be minimized subject to the dynamically generated condition $\abs{\cos\left[\Delta(J_a,J_c)\right]}= 1$ (Fig.~\ref{Fig3}b). The energy reaches minimum $H^{(2)}_0=\overline{W}_{ab}+\overline{W}_{dc}-\frac{\omega_0}{2}\left(1+\frac{\kappa_{ba}^2+\kappa_{dc}^2}{4}+\frac{\left(\kappa_{ba}^2+\kappa_{dc}^2\right)^2}{64}\right)+\mathcal{O}\left(\kappa_{ba}^2+\kappa_{dc}^2\right)^3$ at the point $J_{a,1}=-J_{b,1}\approx \kappa_{ba}/4$ and  $J_{c,1}=-J_{d,1}\approx\kappa_{dc}/4$.  The minimal energy phase-locked stationary solution we interpret as a bound state of two excitons, EM. EM's energy is always lower than the sum of two EPs' energies. The lowest energy of EM is realized for the mirror-symmetric EM (MSEM) configuration with the zero total momentum, i.e. when $\bm k_0\parallel\bm q_a\parallel\bm q_c$ and maximal possible $\kappa_{ba}$. Contrary to the EP case, the stationary solution EM can be reached for a wider range of the parameters $\kappa_{ba}$ and $\kappa_{dc}$. Our numeric investigation shows that the EM solution exists at least for $\abs{\kappa_{ba}}+\abs{\kappa_{dc}}\lesssim 1$ (in Fig.~\ref{Fig3}a $\kappa_{ba}=\kappa_{dc}=0.49$).

{\it Estimation of fluorescent spectra.}
In 2021-2022 a phenomenon of superfluorescence (collective emission of fluorescent light by an ensemble of excited atoms or ions) has been discovered in methylammonium lead iodide beyond 78$K$~\cite{Gundogdu2021Nature_Phot} and in phenethylammonium caesium lead bromide at room-temperature~\cite{Gundogdu2022Nature_Phot}. The superfluorescence effect is manifested by the appearance of a sharp spectral peak (at 2.36$eV$) when the intensity of the pumping light (100$fs$ pulses) exceeds a certain critical value (25.0$\mu Jcm^{-2}$ at 300$K$ and 4.0$\mu Jcm^{-2}$ at 78$K$). The peak appears with a delay of about 10$ps$ (15$ps$ at 78$K$) against the background of a broad (of 0.15$eV$ width) Gaussian fluorescence peak centred at 2.40 $eV$. The Raman~\cite{TDMSKL2016} and theraherz~\cite{ZSHLZMLWC2017, PhysRevLett.145506.2018, C6MH00275G2016} spectroscopic experiments, and the photoluminescence measurements~\cite{WVMEPSGJH2016} of the composite organic-inorganic materials reveal the presence of mixed organic-inorganic LLPM (decay rates 4 -- 6 times smaller than the phonon frequency~\cite{C6MH00275G2016}) with the frequencies ranging from 91$cm^{-1}$ to 225$cm^{-1}$ for various materials. The modes are strongly coupled to charge carriers~\cite{C6MH00275G2016, Yamada2022, Gundogdu2023Adv_Func_Mat, 3sxd-26yj}, the Fr\"ohlich coupling constant $\alpha_{LO}$ is of order 2 (for comparison in GaAs it has the value 0.068~\cite{GaAs1985}), so that the interaction amplitude $D\sim\sqrt{\alpha_{LO}}$ is also large. The latter circumstance makes our theory applicable to perovskites. Indeed, the value of the energy parameter $\omega_0\sim D^2/\omega_{ph}$ in this case can reach significant values and the ratio $\kappa_{ba} \equiv \frac{W_b-W_a}{\omega_0}$ falls into the allowed region $\abs{\kappa_{ba}}<0.38$ for significantly different $W_b$ and $W_a$. The fact that states a and b have different $W_a$ and $W_b$ means that they can be significantly spaced apart in the momentum space to satisfy the second requirement of the theory that a and b are connected by the phonon mode vector $\bm k_0$ and two excitons can form the MSEM geometry as in Fig.~\ref{Fig3}a.

\begin{figure}
\begin{tabular}{lc}
\includegraphics[scale=0.33]{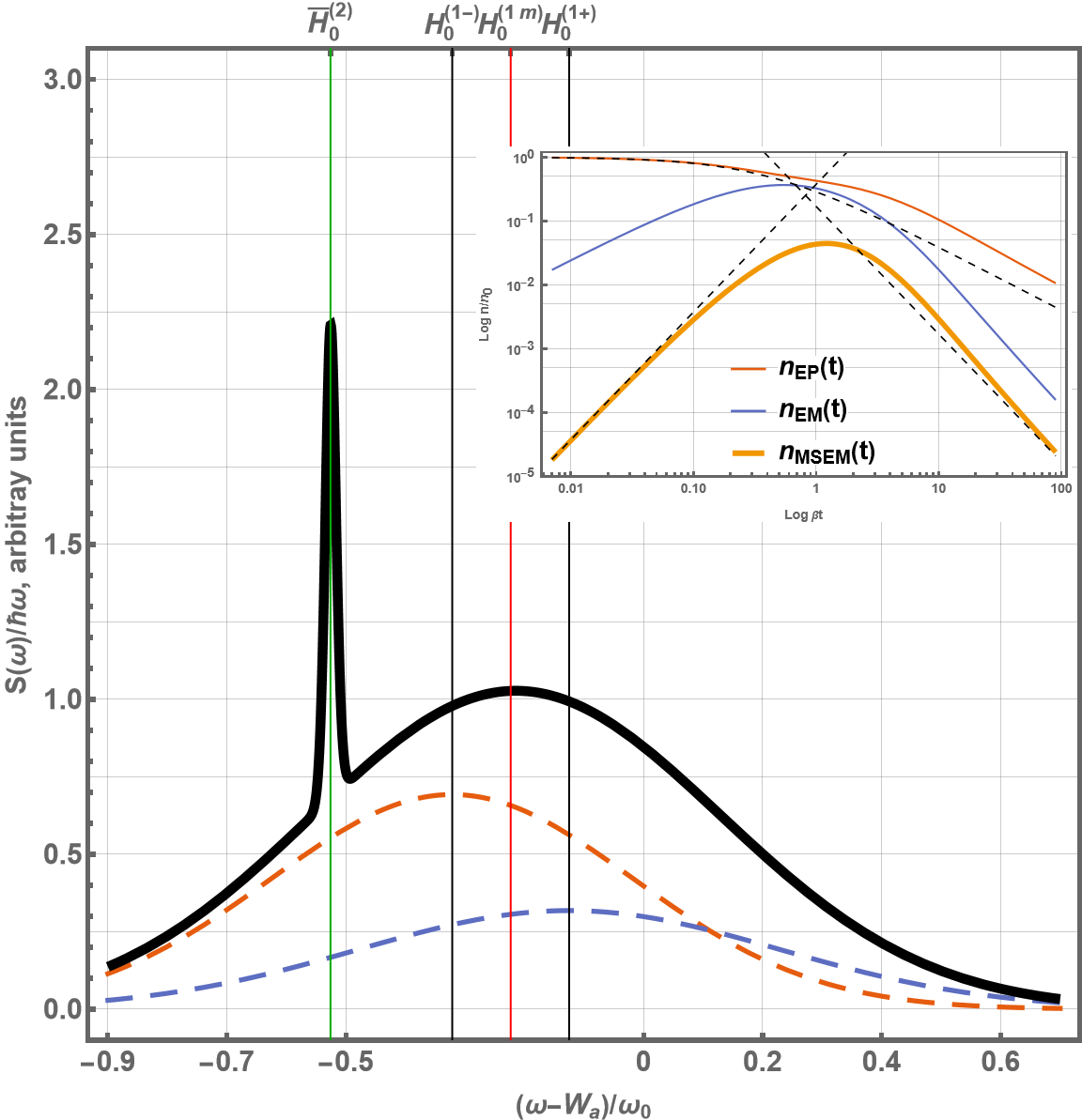}
\end{tabular}
\caption{\small \label{Fig4}The averaged over $\kappa_{ba}$ EP+MSEM spectrum (black curve) plotted vs the frequency $\omega/\omega_0$. For comparison the spectra of single-EP are plotted for the largest $\kappa_{ba}=-0.33$ (orange dashed) and the smallest $\kappa_{ba}=0$ (blue dashed). The maxima are located at $H^{(1\pm)}_0=\left.\left(H^{(1)}_0/ \omega_0-W_a/\omega_0\right)\right|_{\kappa_{ba}=-0.33,\,0}$, and their mean value is denoted by $H^{(1\,m)}_0=(H^{(1+)}_0+H^{(1-)}_0)/2\approx -0.22$. The superfluorescent ($\delta$-)peak is positioned at $\bar{H}_0^{(2)}=\left.H_0^{(2)}/2\omega_0\right|_{\kappa_{ba}=0.49}\sim -0.52$. In the insert: the typical solution of the kinetic equations (eqs.~\ref{dotnbi}--\ref{dotnex}) for the superfluorescent peak intensity (bold orange). The dashed lines shows the asymptotic: $n_{MSEM}\sim \gamma f_1 n_0^2 t^2/2$, $n_{MSEM}\sim n_0 \gamma (\gamma +\beta+f_2) t^{-2}/2f_1\beta(\beta+\gamma-f_2)^2$, and $n_{EP}\sim n_0/(1+f_1n_0t)$ (thin red), the maximum of $n_{MSEM}\sim \gamma f_1 n_0^2  /(\beta+\gamma+f_2)(1+f_1n_0t_0)^2$  is reached at $t_0\sim\sqrt[4]{(\beta+\gamma+f_2)/f_1^2 \gamma (\beta+\gamma-f_2)^2n_0^2}$.   
}
\end{figure}

As an application of our theory, we calculate EP and MSEM fluorescent spectra and compare them with the experimental spectra published in~\cite{Gundogdu2022Nature_Phot}. In particular, we suggest a two-stage mechanism leading to the superfluorescence: the stage of EM formation from two EP and the stage of EM rearrangement into MSEM. The leading terms of the MSEM Hamiltonian coincide with the Dicke Hamiltonian (eq.~\ref{Dicke}), known to describe the superradiance phenomenon~\cite{Leonardi1986DickeMA, Andreev1980CollectiveSE, Protsenko_2015, Mukamel3961997}, which gives the main argument in favour of the suggested mechanism.

By means of the standard technique~\cite{scully1997quantum, Leonardi1986DickeMA} we calculated the EP frequency-resolved stationary fluorescent spectrum, with the initial density matrix entries taken from the stationary solution, $C_i=\sqrt{\frac{1+2J_{i,0}}{2}}e^{-\frac{\i}{2}(W_i+H^{(1)}_0) t}$ and $Q_i=\frac{\sqrt{1-4J_{i,0}^2}}{2} e^{-\i\overline{W}_{ab} t}$. The light-matter interaction term in eq.~(\ref{eq.:H_03}) was taken in the rotation-wave approximation, $\hat{\H}'= \sum\Omega_k \hat{a}^\dag_{\bm k}\hat{a}_{\bm k}+\i F\sum_k (\hat{B}_{\bm k}\hat{a}^\dag_{\bm k}-\hat{B}^\dag_{\bm k}\hat{a}^\dag_{\bm k})$. The natural decay rate into the mode $\Omega_i$ is $\beta_i=4 F^2\beta/[(2\Omega_i-W_i-H^{(1)}_0)^2+4\beta^2]$, where $\beta\sim k_B T$ is the temperature-dependent decoherence rate. The EP fluorescent spectrum $S(\omega)\propto \omega A(\omega-H^{(1)}_0;\kappa_{ba})\delta(\omega-\Omega_a)$ has the Gaussian amplitude $A(\omega-H^{(1)}_0;\kappa_{ba})$ (resulting from averaging over the vibration modes) with the standard deviation of order $\tilde{D}\equiv D(1-4J_{a,0}^2)\sqrt{\frac{1}{8}+\frac{\gamma\beta}{4\omega_{ph}^2}}$ and the prefactor $2\sqrt{\pi} F^2\frac{1-2J_{a,0}}{1+2J_{a,0}}/\tilde{D}$.  The obtained EP spectrum was averaged over all possible directions of the vector $\bm k_0$. Technically, this procedure translates into averaging over $\kappa_{ba}$ within the permitted limits. In our calculations, we took $\kappa_{ba}$ from the range $\abs{\kappa_{ba}}\le 0.33$. As a result, EPs generate a wide fluorescent peak centred at the arithmetic mean of the highest and the lowest EP frequencies (Fig.~\ref{Fig4}). For strong exciton-phonon interaction, $D\sim \omega_0$, the peak position can be approximated by $\approx W_a-0.22 \omega_0$.

For the MSEM model (Fig.~\ref{Fig3}a), when $W_a=W_c$ and $W_b=W_d$, $\bm q_a=-\bm q_c$ and $\bm q_b=-\bm q_d$, the stationary solution of the equations of motion is $C_a=C_c=\sqrt{\frac{1+2J_{a,1}}{2}} e^{-\i \frac{W_a}{2} t -\i \frac{1}{4}H^{(2)}_0 t}$, $C_b=C_d=\sqrt{\frac{1-2J_{a,1}}{2}} e^{-\i\frac{W_b}{2} t  -\i \frac{1}{4}H^{(2)}_0 t}$, and $Q_i=\frac{\sqrt{1-4J_{a,1}^2}}{2} e^{-\i \overline{W}_{ab}t}$. Such high symmetry allows to recast the MSEM Hamiltonian into the form of the Dicke Hamiltonian ($\hat{B}=\hat{B}_a+\hat{B}_c$, $\hat{J}=\hat{J}_a+\hat{J}_c$)
\begin{equation}\label{Dicke}
 \Tilde{\H}_{Dicke}  = \hat{J}H^{(2)}_0/2    +  \Omega_a \hat{a}_1^\dag \hat{a}_1
- \i  F   \big[  \hat{B}^\dag \hat{a}_1 - \hat{B}\hat{a}^\dag_1  \big]/\sqrt{2},
\end{equation}
and the same Hamiltonian for $b$ and $d$. The approximate Hamiltonian (eq.~\ref{Dicke}) follows from the Heisenberg equations written for the steady-state regime when the negligibly small odd combinations, like $\hat{B}_a-\hat{B}_c$, are omitted~\cite{Mukamel3961997, Protsenko_2015}. 

The Hamiltonian (eq.~\ref{Dicke}) commutes with the parity operator $\hat{a}_1^\dag \hat{a}_1+\hat{J}/2$, so that spontaneous decay of one exciton (spontaneous symmetry breaking) causes rapid decay of the other one. This phenomenon is known as enhanced fluorescence and is characterised by the decay rate, twice as fast as the natural one. The light irradiates anisotropically into the mode $\hat{a}_1=(\hat{a}_{\bm q_a}+\hat{a}_{-\bm q_a})/\sqrt{2}$.  The time-dependent spectrum~\cite{Leonardi1986DickeMA} is $S(\omega,t)\propto \pi F^2 \omega \delta(\omega- H^{(2)}_0/2 )\sum_{i=a,b}(1-4J_{i,1}^2)\mathrm{sech}^2\left[2\beta_i (t-\gamma^{-1})\right]$, where $\gamma^{-1}$ now plays a role of the MSEM formation characteristic time. Since in MSEM $\kappa_{ba}=\kappa_{dc}=-0.33$ is fixed (the lowest energy condition), the averaging over $\bm k_0$ results in multiplication of $S(\omega,t)$ on the time-dependent MSEM concentration, $n_{MSEM}(t)$. According to our two-stage model chain of transformations $2EP\leftrightarrows EM\to MSEM\to Light$, we formulate a system of kinetic equations:
\begin{eqnarray}\label{dotnbi}
\dot{n}_{EP}&=&-f_1n_{EP}^2+2f_2n_{EM}.\\
\dot{n}_{EM}&=&f_1n_{EP}^2-f_2n_{EM}-\gamma n_{EM}-\beta n_{EM};\\
\dot{n}_{MSEM}&=&\gamma n_{EM}^2-2\beta n_{MSEM}; \label{dotnex}
\end{eqnarray}
Starting from some non-zero EP concentrations $n_{EP}(0)=n_0$ and $n_{EM}(0)=n_{MSEM}(0)=0$, the sought concentration $n_{MSEM}(t)$ increases quadratically with time (see the insert in Fig.~\ref{Fig4}) reaches the maximal height and decays following the inverse quadratic law. The maximum amplitude is proportional to $n_0$ at some power between 1 and 2 and the time-integrated superfluorescent spectrum generates a sharp peak at the frequency $H^{(2)}_0/2$ (Fig.~\ref{Fig4}). 

Finally, associating the obtained theoretical peaks with their experimental values (Fig.~\ref{Fig4}, compare with Fig.~1e in~\cite{Gundogdu2022Nature_Phot}), we find that $\omega_0\approx0.06eV/(0.52-0.22)=0.2 eV$, so that at $\gamma\sim\omega_{ph}/5$, and $\omega_{ph}=115cm^{-1}=0.014 eV$~\cite{Gundogdu2023Adv_Func_Mat} we get $D\approx0.05eV\approx3.6\omega_{ph}$, which is larger then the room temperature $k_BT_{room}\sim0.026eV$. The obtained value of $D$ is comparable in order of magnitude with the exction binding energy in various types of perovskites $\sim 0.015eV$ -- $0.05eV$~\cite{3sxd-26yj}. Thus the estimate derived from our rough model based on truncated equations of motion yields reasonable values for the energy parameters of the system.

{\it To summarize.} The main result of our theory is that the exciton-phonon interaction leads to the formation of coherent exciton states in solids. Our theory describes a limiting case of the process: the system reaches a stationary state, corresponding to a complete stabilization of the exciton-phonon field. The nonlinear nature of the theory allows for the use of the soliton theory language~\cite{Aklan_2021}. Indeed, the infinitely long-living (up to radiative decay) solutions of the PE and EM obtained in this work can be viewed as three-component multidimensional solitary waves. However, a detailed study of this type of solution requires analysis of its stability beyond the truncated equations of motion, which can become possible by implementing advanced numerical methods. The specific properties of organic-based perovskites made the application of our model to the discussion of the room-temperature superfluorescence phenomenon possible. Recently, two approaches based on known soliton models (like the extended non-linear Schr\"odinger equation) have been independently proposed in experimental~\cite{GSoliton} and theoretical~\cite{gladkij2025stabilityquantumcoherentsuperradiant} works for describing the superfluorescence phenomenon in perovskites. In the works~\cite{Gundogdu2022Nature_Phot} and~\cite{Fainberg_Osipov24JCP}, the phenomenon of superfluorescence was considered as a result of emission from macroscopic coherent states of polarons. The exciton molecules introduced in this work are coherent pairs of excitons with a total zero momentum, and there are no theoretical objections to the creation of a network of coherent pairs that can also form a macroscopic coherence. 

The presented dynamic approach for EP description paves the way for further generalisations by incorporating the thermal bath stochastic terms, Coulomb interaction terms, and other factors to describe the dynamics of EP transformations to a coherent macroscopic state of excitons.

{\it Acknowledgement.} We thank Kenan Gundogdu for stimulating our interest to the problem and useful discussions. We thank him for paying our attention to Ref.~\cite{GSoliton}.


\begin{thebibliography}{43}%
\makeatletter
\providecommand \@ifxundefined [1]{%
 \@ifx{#1\undefined}
}%
\providecommand \@ifnum [1]{%
 \ifnum #1\expandafter \@firstoftwo
 \else \expandafter \@secondoftwo
 \fi
}%
\providecommand \@ifx [1]{%
 \ifx #1\expandafter \@firstoftwo
 \else \expandafter \@secondoftwo
 \fi
}%
\providecommand \natexlab [1]{#1}%
\providecommand \enquote  [1]{``#1''}%
\providecommand \bibnamefont  [1]{#1}%
\providecommand \bibfnamefont [1]{#1}%
\providecommand \citenamefont [1]{#1}%
\providecommand \href@noop [0]{\@secondoftwo}%
\providecommand \href [0]{\begingroup \@sanitize@url \@href}%
\providecommand \@href[1]{\@@startlink{#1}\@@href}%
\providecommand \@@href[1]{\endgroup#1\@@endlink}%
\providecommand \@sanitize@url [0]{\catcode `\\12\catcode `\$12\catcode
  `\&12\catcode `\#12\catcode `\^12\catcode `\_12\catcode `\%12\relax}%
\providecommand \@@startlink[1]{}%
\providecommand \@@endlink[0]{}%
\providecommand \url  [0]{\begingroup\@sanitize@url \@url }%
\providecommand \@url [1]{\endgroup\@href {#1}{\urlprefix }}%
\providecommand \urlprefix  [0]{URL }%
\providecommand \Eprint [0]{\href }%
\providecommand \doibase [0]{https://doi.org/}%
\providecommand \selectlanguage [0]{\@gobble}%
\providecommand \bibinfo  [0]{\@secondoftwo}%
\providecommand \bibfield  [0]{\@secondoftwo}%
\providecommand \translation [1]{[#1]}%
\providecommand \BibitemOpen [0]{}%
\providecommand \bibitemStop [0]{}%
\providecommand \bibitemNoStop [0]{.\EOS\space}%
\providecommand \EOS [0]{\spacefactor3000\relax}%
\providecommand \BibitemShut  [1]{\csname bibitem#1\endcsname}%
\let\auto@bib@innerbib\@empty
\bibitem [{\citenamefont {Pyshkin}(2018)}]{Pyshkin_2018}%
  \BibitemOpen
  \bibfield  {author} {\bibinfo {author} {\bibfnamefont {S.~L.}\ \bibnamefont
  {Pyshkin}},\ }\href {https://doi.org/10.5772/66232} {\emph {\bibinfo {title}
  {Excitons}}}\ (\bibinfo  {publisher} {IntechOpen},\ \bibinfo {address}
  {Rijeka},\ \bibinfo {year} {2018})\BibitemShut {NoStop}%
\bibitem [{\citenamefont {Anantharaman}\ \emph {et~al.}(2021)\citenamefont
  {Anantharaman}, \citenamefont {Jo},\ and\ \citenamefont
  {Jariwala}}]{ExcitonReview2021}%
  \BibitemOpen
  \bibfield  {author} {\bibinfo {author} {\bibfnamefont {S.~B.}\ \bibnamefont
  {Anantharaman}}, \bibinfo {author} {\bibfnamefont {K.}~\bibnamefont {Jo}},\
  and\ \bibinfo {author} {\bibfnamefont {D.}~\bibnamefont {Jariwala}},\
  }\bibfield  {title} {\bibinfo {title} {Exciton–photonics: From fundamental
  science to applications},\ }\href {https://doi.org/10.1021/acsnano.1c02204}
  {\bibfield  {journal} {\bibinfo  {journal} {ACS Nano}\ }\textbf {\bibinfo
  {volume} {15}},\ \bibinfo {pages} {12628} (\bibinfo {year}
  {2021})}\BibitemShut {NoStop}%
\bibitem [{\citenamefont {Dhar}(2025)}]{Subhabrata2025}%
  \BibitemOpen
  \bibfield  {author} {\bibinfo {author} {\bibfnamefont {S.}~\bibnamefont
  {Dhar}},\ }\href {https://doi.org/10.1007/978-3-031-87278-5} {\emph {\bibinfo
  {title} {Excitons in Semiconductors}}}\ (\bibinfo  {publisher} {Springer
  Cham},\ \bibinfo {address} {Switzerland},\ \bibinfo {year}
  {2025})\BibitemShut {NoStop}%
\bibitem [{\citenamefont {Emin}(2012)}]{Emin_2012}%
  \BibitemOpen
  \bibfield  {author} {\bibinfo {author} {\bibfnamefont {D.}~\bibnamefont
  {Emin}},\ }\href@noop {} {\emph {\bibinfo {title} {Polarons}}}\ (\bibinfo
  {publisher} {Cambridge University Press},\ \bibinfo {year}
  {2012})\BibitemShut {NoStop}%
\bibitem [{\citenamefont {Schmidt}(1969)}]{Schmidt69}%
  \BibitemOpen
  \bibfield  {author} {\bibinfo {author} {\bibfnamefont {P.~P.}\ \bibnamefont
  {Schmidt}},\ }\bibfield  {title} {\bibinfo {title} {The interaction between
  mott-wannier excitons in a vibrating lattice},\ }\href@noop {} {\bibfield
  {journal} {\bibinfo  {journal} {J. Phys. C: Solid State Phys.}\ }\textbf
  {\bibinfo {volume} {2}},\ \bibinfo {pages} {785} (\bibinfo {year}
  {1969})}\BibitemShut {NoStop}%
\bibitem [{\citenamefont {Iadonisi}\ and\ \citenamefont
  {Bassani}(1983)}]{IB1983}%
  \BibitemOpen
  \bibfield  {author} {\bibinfo {author} {\bibfnamefont {G.}~\bibnamefont
  {Iadonisi}}\ and\ \bibinfo {author} {\bibfnamefont {F.}~\bibnamefont
  {Bassani}},\ }\bibfield  {title} {\bibinfo {title} {Excitonic polaron states
  and optical transitions},\ }\href
  {https://doi.org/https://doi.org/10.1007/BF02460231} {\bibfield  {journal}
  {\bibinfo  {journal} {Il Nuovo Cimento D}\ }\textbf {\bibinfo {volume} {2}},\
  \bibinfo {pages} {1541–1560} (\bibinfo {year} {1983})}\BibitemShut
  {NoStop}%
\bibitem [{\citenamefont {Singh}\ and\ \citenamefont
  {Matsui}(1987)}]{SinghMatsui1987}%
  \BibitemOpen
  \bibfield  {author} {\bibinfo {author} {\bibfnamefont {J.}~\bibnamefont
  {Singh}}\ and\ \bibinfo {author} {\bibfnamefont {A.}~\bibnamefont {Matsui}},\
  }\bibfield  {title} {\bibinfo {title} {Free, quasifree, momentarily trapped,
  and self-trapped exciton states in molecular crystals},\ }\href
  {https://doi.org/10.1103/PhysRevB.36.6094} {\bibfield  {journal} {\bibinfo
  {journal} {Phys. Rev. B}\ }\textbf {\bibinfo {volume} {36}},\ \bibinfo
  {pages} {6094} (\bibinfo {year} {1987})}\BibitemShut {NoStop}%
\bibitem [{\citenamefont {Dai}\ \emph {et~al.}(2024)\citenamefont {Dai},
  \citenamefont {Lian}, \citenamefont {Lafuente-Bartolome},\ and\ \citenamefont
  {Giustino}}]{ZhChJF2024}%
  \BibitemOpen
  \bibfield  {author} {\bibinfo {author} {\bibfnamefont {Z.}~\bibnamefont
  {Dai}}, \bibinfo {author} {\bibfnamefont {C.}~\bibnamefont {Lian}}, \bibinfo
  {author} {\bibfnamefont {J.}~\bibnamefont {Lafuente-Bartolome}},\ and\
  \bibinfo {author} {\bibfnamefont {F.}~\bibnamefont {Giustino}},\ }\bibfield
  {title} {\bibinfo {title} {Theory of excitonic polarons: From models to
  first-principles calculations},\ }\href
  {https://doi.org/10.1103/PhysRevB.109.045202} {\bibfield  {journal} {\bibinfo
   {journal} {Phys. Rev. B}\ }\textbf {\bibinfo {volume} {109}},\ \bibinfo
  {pages} {045202} (\bibinfo {year} {2024})}\BibitemShut {NoStop}%
\bibitem [{\citenamefont {Singh}\ \emph {et~al.}(1996)\citenamefont {Singh},
  \citenamefont {Birkedal}, \citenamefont {Lyssenko},\ and\ \citenamefont
  {Hvam}}]{Singh2Dbiexciton1996}%
  \BibitemOpen
  \bibfield  {author} {\bibinfo {author} {\bibfnamefont {J.}~\bibnamefont
  {Singh}}, \bibinfo {author} {\bibfnamefont {D.}~\bibnamefont {Birkedal}},
  \bibinfo {author} {\bibfnamefont {V.~G.}\ \bibnamefont {Lyssenko}},\ and\
  \bibinfo {author} {\bibfnamefont {J.~M.}\ \bibnamefont {Hvam}},\ }\bibfield
  {title} {\bibinfo {title} {Binding energy of two-dimensional biexcitons},\
  }\href {https://doi.org/10.1103/PhysRevB.53.15909} {\bibfield  {journal}
  {\bibinfo  {journal} {Phys. Rev. B}\ }\textbf {\bibinfo {volume} {53}},\
  \bibinfo {pages} {15909} (\bibinfo {year} {1996})}\BibitemShut {NoStop}%
\bibitem [{\citenamefont {Li}\ \emph {et~al.}(2020)\citenamefont {Li},
  \citenamefont {Ma}, \citenamefont {Wang}, \citenamefont {Fang}, \citenamefont
  {Luo},\ and\ \citenamefont {Li}}]{LiMaWangFangLuoLi+2020+2001+2006}%
  \BibitemOpen
  \bibfield  {author} {\bibinfo {author} {\bibfnamefont {W.}~\bibnamefont
  {Li}}, \bibinfo {author} {\bibfnamefont {J.}~\bibnamefont {Ma}}, \bibinfo
  {author} {\bibfnamefont {H.}~\bibnamefont {Wang}}, \bibinfo {author}
  {\bibfnamefont {C.}~\bibnamefont {Fang}}, \bibinfo {author} {\bibfnamefont
  {H.}~\bibnamefont {Luo}},\ and\ \bibinfo {author} {\bibfnamefont
  {D.}~\bibnamefont {Li}},\ }\bibfield  {title} {\bibinfo {title} {Biexcitons
  in 2d (iso-ba)2pbi4 perovskite crystals},\ }\href
  {https://doi.org/doi:10.1515/nanoph-2019-0528} {\bibfield  {journal}
  {\bibinfo  {journal} {Nanophotonics}\ }\textbf {\bibinfo {volume} {9}},\
  \bibinfo {pages} {2001} (\bibinfo {year} {2020})}\BibitemShut {NoStop}%
\bibitem [{\citenamefont {Elliott}(1957)}]{E1957}%
  \BibitemOpen
  \bibfield  {author} {\bibinfo {author} {\bibfnamefont {R.~J.}\ \bibnamefont
  {Elliott}},\ }\bibfield  {title} {\bibinfo {title} {Intensity of optical
  absorption by excitons},\ }\href {https://doi.org/10.1103/PhysRev.108.1384}
  {\bibfield  {journal} {\bibinfo  {journal} {Phys. Rev.}\ }\textbf {\bibinfo
  {volume} {108}},\ \bibinfo {pages} {1384} (\bibinfo {year}
  {1957})}\BibitemShut {NoStop}%
\bibitem [{\citenamefont {Singh}(1994)}]{Singh1994}%
  \BibitemOpen
  \bibfield  {author} {\bibinfo {author} {\bibfnamefont {J.}~\bibnamefont
  {Singh}},\ }\bibinfo {title} {Excitation energy transfer processes in
  condensed matter}\ (\bibinfo  {publisher} {Springer},\ \bibinfo {address}
  {Boston, MA},\ \bibinfo {year} {1994})\BibitemShut {NoStop}%
\bibitem [{\citenamefont {Fr\"ohlich}(1954)}]{Frohlich01071954}%
  \BibitemOpen
  \bibfield  {author} {\bibinfo {author} {\bibfnamefont {H.}~\bibnamefont
  {Fr\"ohlich}},\ }\bibfield  {title} {\bibinfo {title} {Electrons in lattice
  fields},\ }\href {https://doi.org/10.1080/00018735400101213} {\bibfield
  {journal} {\bibinfo  {journal} {Advances in Physics}\ }\textbf {\bibinfo
  {volume} {3}},\ \bibinfo {pages} {325} (\bibinfo {year} {1954})}\BibitemShut
  {NoStop}%
\bibitem [{\citenamefont {Antonius}\ and\ \citenamefont
  {Louie}(2022)}]{AL2022}%
  \BibitemOpen
  \bibfield  {author} {\bibinfo {author} {\bibfnamefont {G.}~\bibnamefont
  {Antonius}}\ and\ \bibinfo {author} {\bibfnamefont {S.~G.}\ \bibnamefont
  {Louie}},\ }\bibfield  {title} {\bibinfo {title} {Theory of exciton-phonon
  coupling},\ }\href {https://doi.org/10.1103/PhysRevB.105.085111} {\bibfield
  {journal} {\bibinfo  {journal} {Phys. Rev. B}\ }\textbf {\bibinfo {volume}
  {105}},\ \bibinfo {pages} {085111} (\bibinfo {year} {2022})}\BibitemShut
  {NoStop}%
\bibitem [{\citenamefont {Combescot}\ \emph {et~al.}(2007)\citenamefont
  {Combescot}, \citenamefont {Betbeder-Matibet},\ and\ \citenamefont
  {Dubin}}]{CBD2007}%
  \BibitemOpen
  \bibfield  {author} {\bibinfo {author} {\bibfnamefont {M.}~\bibnamefont
  {Combescot}}, \bibinfo {author} {\bibfnamefont {O.}~\bibnamefont
  {Betbeder-Matibet}},\ and\ \bibinfo {author} {\bibfnamefont {F.}~\bibnamefont
  {Dubin}},\ }\bibfield  {title} {\bibinfo {title} {Mixture of composite-boson
  molecules and the pauli exclusion principle},\ }\href
  {https://doi.org/10.1103/PhysRevA.76.033601} {\bibfield  {journal} {\bibinfo
  {journal} {Phys. Rev. A}\ }\textbf {\bibinfo {volume} {76}},\ \bibinfo
  {pages} {033601} (\bibinfo {year} {2007})}\BibitemShut {NoStop}%
\bibitem [{\citenamefont {Davydov}\ and\ \citenamefont
  {Kislukha}(1973)}]{Davydov1973}%
  \BibitemOpen
  \bibfield  {author} {\bibinfo {author} {\bibfnamefont {A.~S.}\ \bibnamefont
  {Davydov}}\ and\ \bibinfo {author} {\bibfnamefont {N.~I.}\ \bibnamefont
  {Kislukha}},\ }\bibfield  {title} {\bibinfo {title} {Solitary excitons in
  one-dimensional molecular chains},\ }\href
  {https://doi.org/https://doi.org/10.1002/pssb.2220590212} {\bibfield
  {journal} {\bibinfo  {journal} {Physica Status Solidi (b)}\ }\textbf
  {\bibinfo {volume} {59}},\ \bibinfo {pages} {465} (\bibinfo {year}
  {1973})}\BibitemShut {NoStop}%
\bibitem [{\citenamefont {Meyer}\ \emph {et~al.}(1990)\citenamefont {Meyer},
  \citenamefont {Manthe},\ and\ \citenamefont {Cederbaum}}]{MEYER199073}%
  \BibitemOpen
  \bibfield  {author} {\bibinfo {author} {\bibfnamefont {H.-D.}\ \bibnamefont
  {Meyer}}, \bibinfo {author} {\bibfnamefont {U.}~\bibnamefont {Manthe}},\ and\
  \bibinfo {author} {\bibfnamefont {L.}~\bibnamefont {Cederbaum}},\ }\bibfield
  {title} {\bibinfo {title} {The multi-configurational time-dependent hartree
  approach},\ }\href
  {https://doi.org/https://doi.org/10.1016/0009-2614(90)87014-I} {\bibfield
  {journal} {\bibinfo  {journal} {Chemical Physics Letters}\ }\textbf {\bibinfo
  {volume} {165}},\ \bibinfo {pages} {73} (\bibinfo {year} {1990})}\BibitemShut
  {NoStop}%
\bibitem [{\citenamefont {Miller}(2002)}]{Miller2002}%
  \BibitemOpen
  \bibfield  {author} {\bibinfo {author} {\bibfnamefont {W.~H.}\ \bibnamefont
  {Miller}},\ }\bibfield  {title} {\bibinfo {title} {On the relation between
  the semiclassical initial value representation and an exact quantum expansion
  in time-dependent coherent states},\ }\href
  {https://doi.org/10.1021/jp020500} {\bibfield  {journal} {\bibinfo  {journal}
  {Journal of Physical Chemistry B}\ }\textbf {\bibinfo {volume} {106}},\
  \bibinfo {pages} {8132} (\bibinfo {year} {2002})}\BibitemShut {NoStop}%
\bibitem [{\citenamefont {Artacho}\ and\ \citenamefont
  {O'Regan}(2017)}]{AR2017}%
  \BibitemOpen
  \bibfield  {author} {\bibinfo {author} {\bibfnamefont {E.}~\bibnamefont
  {Artacho}}\ and\ \bibinfo {author} {\bibfnamefont {D.}~\bibnamefont
  {O'Regan}},\ }\bibfield  {title} {\bibinfo {title} {Quantum mechanics in an
  evolving hilbert space},\ }\href {https://doi.org/10.1021/jp020500+}
  {\bibfield  {journal} {\bibinfo  {journal} {Physical Review B}\ }\textbf
  {\bibinfo {volume} {95}},\ \bibinfo {pages} {115155} (\bibinfo {year}
  {2017})}\BibitemShut {NoStop}%
\bibitem [{\citenamefont {Zhao}\ \emph {et~al.}(2022)\citenamefont {Zhao},
  \citenamefont {Sun}, \citenamefont {Chen},\ and\ \citenamefont
  {Gelin}}]{Zhao2022}%
  \BibitemOpen
  \bibfield  {author} {\bibinfo {author} {\bibfnamefont {Y.}~\bibnamefont
  {Zhao}}, \bibinfo {author} {\bibfnamefont {K.}~\bibnamefont {Sun}}, \bibinfo
  {author} {\bibfnamefont {L.}~\bibnamefont {Chen}},\ and\ \bibinfo {author}
  {\bibfnamefont {M.}~\bibnamefont {Gelin}},\ }\bibfield  {title} {\bibinfo
  {title} {The hierarchy of davydov's ansätze and its applications},\ }\href
  {https://doi.org/https://doi.org/10.1002/wcms.1589} {\bibfield  {journal}
  {\bibinfo  {journal} {WIREs Computational Molecular Science}\ }\textbf
  {\bibinfo {volume} {12}},\ \bibinfo {pages} {e1589} (\bibinfo {year}
  {2022})}\BibitemShut {NoStop}%
\bibitem [{\citenamefont {Osipov}\ and\ \citenamefont
  {B.Fainberg}(2023)}]{Osipov_Fainberg23PRB}%
  \BibitemOpen
  \bibfield  {author} {\bibinfo {author} {\bibfnamefont {V.}~\bibnamefont
  {Osipov}}\ and\ \bibinfo {author} {\bibnamefont {B.Fainberg}},\ }\bibfield
  {title} {\bibinfo {title} {Hartree method for molecular polaritons},\
  }\href@noop {} {\bibfield  {journal} {\bibinfo  {journal} {Phys. Rev. B}\
  }\textbf {\bibinfo {volume} {107}},\ \bibinfo {pages} {075404} (\bibinfo
  {year} {2023})}\BibitemShut {NoStop}%
\bibitem [{\citenamefont {Aklan}\ \emph {et~al.}(2021)\citenamefont {Aklan},
  \citenamefont {Ishak},\ and\ \citenamefont {Umarov}}]{Aklan_2021}%
  \BibitemOpen
  \bibfield  {author} {\bibinfo {author} {\bibfnamefont {N.~A.~B.}\
  \bibnamefont {Aklan}}, \bibinfo {author} {\bibfnamefont {N.~A.~S.}\
  \bibnamefont {Ishak}},\ and\ \bibinfo {author} {\bibfnamefont {B.~A.}\
  \bibnamefont {Umarov}},\ }\bibfield  {title} {\bibinfo {title} {Vector
  soliton in coupled nonlinear schrödinger equation},\ }\href
  {https://doi.org/10.1088/1742-6596/1988/1/012015} {\bibfield  {journal}
  {\bibinfo  {journal} {Journal of Physics: Conference Series}\ }\textbf
  {\bibinfo {volume} {1988}},\ \bibinfo {pages} {012015} (\bibinfo {year}
  {2021})}\BibitemShut {NoStop}%
\bibitem [{\citenamefont {Toyozawa}(1959)}]{Toyozawa59}%
  \BibitemOpen
  \bibfield  {author} {\bibinfo {author} {\bibfnamefont {Y.}~\bibnamefont
  {Toyozawa}},\ }\bibfield  {title} {\bibinfo {title} {On the dynamical
  behavior of an exciton},\ }\href@noop {} {\bibfield  {journal} {\bibinfo
  {journal} {Progr. Theor. Phys. Suppl.}\ }\textbf {\bibinfo {volume} {12}},\
  \bibinfo {pages} {111} (\bibinfo {year} {1959})}\BibitemShut {NoStop}%
\bibitem [{\citenamefont {Fainberg}\ and\ \citenamefont
  {Osipov}(2024)}]{Fainberg_Osipov24JCP}%
  \BibitemOpen
  \bibfield  {author} {\bibinfo {author} {\bibfnamefont {B.~D.}\ \bibnamefont
  {Fainberg}}\ and\ \bibinfo {author} {\bibfnamefont {V.~A.}\ \bibnamefont
  {Osipov}},\ }\bibfield  {title} {\bibinfo {title} {Theory of high-temperature
  superfluorescence in hybrid perovskite thin films},\ }\href
  {https://doi.org/10.1063/5.0226221} {\bibfield  {journal} {\bibinfo
  {journal} {Journal of Chemical Physics}\ }\textbf {\bibinfo {volume} {161}},\
  \bibinfo {pages} {114705} (\bibinfo {year} {2024})}\BibitemShut {NoStop}%
\bibitem [{Note1()}]{Note1}%
  \BibitemOpen
  \bibinfo {note} {\begin {multline*} D(k)=\omega _{ph}\protect \sqrt {\protect
  \frac {\alpha _{LO}}{V\protect \sqrt {2m^*_e \omega _{ph}}}}\protect \tmspace
  +\thickmuskip {.2777em}\protect \frac {4\pi } {k}\\ \left [ \left (1+\left
  (\protect \frac {\mu _e a_0 k}{2}\right )^2\right )^{-\xi } -\left (1+\left
  (\protect \frac {m^*_h \mu _e a_0 k}{2m^*_e}\right )^2\right )^{-\xi } \right
  ], \end {multline*} where $a_0$ is the exciton Bohr radius, $\mu _e$ is the
  ratio of the electron effective mass and the exciton reduced mass, $\mu
  _e=m_h^*/M_{ex}$, $\alpha _{LO}$ is the coupling constant, and $V$ is the
  crystal volume. The exponent $\xi =2$ for the bulk material~\cite
  {Toyozawa59} and changes to $\xi =3/2$ for the geometry of thin films~\cite
  {Fainberg_Osipov24JCP}.}\BibitemShut {Stop}%
\bibitem [{\citenamefont {Findik}\ \emph {et~al.}(2021)\citenamefont {Findik},
  \citenamefont {Biliroglu}, \citenamefont {Seyitliyev}, \citenamefont
  {Mendes}, \citenamefont {Barrette}, \citenamefont {Ardekani}, \citenamefont
  {Lei}, \citenamefont {Dong}, \citenamefont {So},\ and\ \citenamefont
  {Gundogdu}}]{Gundogdu2021Nature_Phot}%
  \BibitemOpen
  \bibfield  {author} {\bibinfo {author} {\bibfnamefont {G.}~\bibnamefont
  {Findik}}, \bibinfo {author} {\bibfnamefont {M.}~\bibnamefont {Biliroglu}},
  \bibinfo {author} {\bibfnamefont {D.}~\bibnamefont {Seyitliyev}}, \bibinfo
  {author} {\bibfnamefont {J.}~\bibnamefont {Mendes}}, \bibinfo {author}
  {\bibfnamefont {A.}~\bibnamefont {Barrette}}, \bibinfo {author}
  {\bibfnamefont {H.}~\bibnamefont {Ardekani}}, \bibinfo {author}
  {\bibfnamefont {L.}~\bibnamefont {Lei}}, \bibinfo {author} {\bibfnamefont
  {Q.}~\bibnamefont {Dong}}, \bibinfo {author} {\bibfnamefont {F.}~\bibnamefont
  {So}},\ and\ \bibinfo {author} {\bibfnamefont {K.}~\bibnamefont {Gundogdu}},\
  }\bibfield  {title} {\bibinfo {title} {High-temperature superfluorescence in
  methyl ammonium lead iodide},\ }\href
  {https://doi.org/10.1038/s41566-021-00830-x} {\bibfield  {journal} {\bibinfo
  {journal} {Nature Photonics}\ }\textbf {\bibinfo {volume} {15}},\ \bibinfo
  {pages} {676 } (\bibinfo {year} {2021})}\BibitemShut {NoStop}%
\bibitem [{\citenamefont {Biliroglu}\ \emph {et~al.}(2022)\citenamefont
  {Biliroglu}, \citenamefont {Findik}, \citenamefont {Mendes}, \citenamefont
  {Seyitliyev}, \citenamefont {Lei}, \citenamefont {Dong}, \citenamefont
  {Mehta}, \citenamefont {Temnov}, \citenamefont {So},\ and\ \citenamefont
  {Gundogdu}}]{Gundogdu2022Nature_Phot}%
  \BibitemOpen
  \bibfield  {author} {\bibinfo {author} {\bibfnamefont {M.}~\bibnamefont
  {Biliroglu}}, \bibinfo {author} {\bibfnamefont {G.}~\bibnamefont {Findik}},
  \bibinfo {author} {\bibfnamefont {J.}~\bibnamefont {Mendes}}, \bibinfo
  {author} {\bibfnamefont {D.}~\bibnamefont {Seyitliyev}}, \bibinfo {author}
  {\bibfnamefont {L.}~\bibnamefont {Lei}}, \bibinfo {author} {\bibfnamefont
  {Q.}~\bibnamefont {Dong}}, \bibinfo {author} {\bibfnamefont {Y.}~\bibnamefont
  {Mehta}}, \bibinfo {author} {\bibfnamefont {V.~V.}\ \bibnamefont {Temnov}},
  \bibinfo {author} {\bibfnamefont {F.}~\bibnamefont {So}},\ and\ \bibinfo
  {author} {\bibfnamefont {K.}~\bibnamefont {Gundogdu}},\ }\bibfield  {title}
  {\bibinfo {title} {Room-temperature superfluorescence in hybrid perovskites
  and its origins},\ }\href@noop {} {\bibfield  {journal} {\bibinfo  {journal}
  {Nature Photonics}\ }\textbf {\bibinfo {volume} {16}},\ \bibinfo {pages}
  {324} (\bibinfo {year} {2022})}\BibitemShut {NoStop}%
\bibitem [{\citenamefont {Tilchin}\ \emph {et~al.}(2016)\citenamefont
  {Tilchin}, \citenamefont {Dirin}, \citenamefont {Maikov}, \citenamefont
  {Sashchiuk}, \citenamefont {Kovalenko},\ and\ \citenamefont
  {Lifshitz}}]{TDMSKL2016}%
  \BibitemOpen
  \bibfield  {author} {\bibinfo {author} {\bibfnamefont {J.}~\bibnamefont
  {Tilchin}}, \bibinfo {author} {\bibfnamefont {D.~N.}\ \bibnamefont {Dirin}},
  \bibinfo {author} {\bibfnamefont {G.~I.}\ \bibnamefont {Maikov}}, \bibinfo
  {author} {\bibfnamefont {A.}~\bibnamefont {Sashchiuk}}, \bibinfo {author}
  {\bibfnamefont {M.~V.}\ \bibnamefont {Kovalenko}},\ and\ \bibinfo {author}
  {\bibfnamefont {E.}~\bibnamefont {Lifshitz}},\ }\bibfield  {title} {\bibinfo
  {title} {Hydrogen-like wannier–mott excitons in single crystal of
  methylammonium lead bromide perovskite},\ }\href
  {https://doi.org/10.1021/acsnano.6b02734} {\bibfield  {journal} {\bibinfo
  {journal} {ACS Nano}\ }\textbf {\bibinfo {volume} {10}},\ \bibinfo {pages}
  {6363} (\bibinfo {year} {2016})},\ \bibinfo {note} {pMID:
  27249335}\BibitemShut {NoStop}%
\bibitem [{\citenamefont {Zhao}\ \emph {et~al.}(2017)\citenamefont {Zhao},
  \citenamefont {Skelton}, \citenamefont {Hu}, \citenamefont {La-o vorakiat},
  \citenamefont {Zhu}, \citenamefont {Marcus}, \citenamefont {Michel-Beyerle},
  \citenamefont {Lam}, \citenamefont {Walsh},\ and\ \citenamefont
  {Chia}}]{ZSHLZMLWC2017}%
  \BibitemOpen
  \bibfield  {author} {\bibinfo {author} {\bibfnamefont {D.}~\bibnamefont
  {Zhao}}, \bibinfo {author} {\bibfnamefont {J.~M.}\ \bibnamefont {Skelton}},
  \bibinfo {author} {\bibfnamefont {H.}~\bibnamefont {Hu}}, \bibinfo {author}
  {\bibfnamefont {C.}~\bibnamefont {La-o vorakiat}}, \bibinfo {author}
  {\bibfnamefont {J.-X.}\ \bibnamefont {Zhu}}, \bibinfo {author} {\bibfnamefont
  {R.~A.}\ \bibnamefont {Marcus}}, \bibinfo {author} {\bibfnamefont {M.-E.}\
  \bibnamefont {Michel-Beyerle}}, \bibinfo {author} {\bibfnamefont {Y.~M.}\
  \bibnamefont {Lam}}, \bibinfo {author} {\bibfnamefont {A.}~\bibnamefont
  {Walsh}},\ and\ \bibinfo {author} {\bibfnamefont {E.~E.~M.}\ \bibnamefont
  {Chia}},\ }\bibfield  {title} {\bibinfo {title} {Low-frequency optical phonon
  modes and carrier mobility in the halide perovskite ch3nh3pbbr3 using
  terahertz time-domain spectroscopy},\ }\href
  {https://doi.org/10.1063/1.4993524} {\bibfield  {journal} {\bibinfo
  {journal} {Applied Physics Letters}\ }\textbf {\bibinfo {volume} {111}},\
  \bibinfo {pages} {201903} (\bibinfo {year} {2017})}\BibitemShut {NoStop}%
\bibitem [{\citenamefont {Nagai}\ \emph {et~al.}(2018)\citenamefont {Nagai},
  \citenamefont {Tomioka}, \citenamefont {Ashida}, \citenamefont {Hoyano},
  \citenamefont {Akashi}, \citenamefont {Yamada}, \citenamefont {Aharen},\ and\
  \citenamefont {Kanemitsu}}]{PhysRevLett.145506.2018}%
  \BibitemOpen
  \bibfield  {author} {\bibinfo {author} {\bibfnamefont {M.}~\bibnamefont
  {Nagai}}, \bibinfo {author} {\bibfnamefont {T.}~\bibnamefont {Tomioka}},
  \bibinfo {author} {\bibfnamefont {M.}~\bibnamefont {Ashida}}, \bibinfo
  {author} {\bibfnamefont {M.}~\bibnamefont {Hoyano}}, \bibinfo {author}
  {\bibfnamefont {R.}~\bibnamefont {Akashi}}, \bibinfo {author} {\bibfnamefont
  {Y.}~\bibnamefont {Yamada}}, \bibinfo {author} {\bibfnamefont
  {T.}~\bibnamefont {Aharen}},\ and\ \bibinfo {author} {\bibfnamefont
  {Y.}~\bibnamefont {Kanemitsu}},\ }\bibfield  {title} {\bibinfo {title}
  {Longitudinal optical phonons modified by organic molecular cation motions in
  organic-inorganic hybrid perovskites},\ }\href
  {https://doi.org/10.1103/PhysRevLett.121.145506} {\bibfield  {journal}
  {\bibinfo  {journal} {Phys. Rev. Lett.}\ }\textbf {\bibinfo {volume} {121}},\
  \bibinfo {pages} {145506} (\bibinfo {year} {2018})}\BibitemShut {NoStop}%
\bibitem [{\citenamefont {Sendner}\ \emph {et~al.}(2016)\citenamefont
  {Sendner}, \citenamefont {Nayak}, \citenamefont {Egger}, \citenamefont
  {Beck}, \citenamefont {Müller}, \citenamefont {Epding}, \citenamefont
  {Kowalsky}, \citenamefont {Kronik}, \citenamefont {Snaith}, \citenamefont
  {Pucci},\ and\ \citenamefont {Lovrin\v{c}i\'c}}]{C6MH00275G2016}%
  \BibitemOpen
  \bibfield  {author} {\bibinfo {author} {\bibfnamefont {M.}~\bibnamefont
  {Sendner}}, \bibinfo {author} {\bibfnamefont {P.~K.}\ \bibnamefont {Nayak}},
  \bibinfo {author} {\bibfnamefont {D.~A.}\ \bibnamefont {Egger}}, \bibinfo
  {author} {\bibfnamefont {S.}~\bibnamefont {Beck}}, \bibinfo {author}
  {\bibfnamefont {C.}~\bibnamefont {Müller}}, \bibinfo {author} {\bibfnamefont
  {B.}~\bibnamefont {Epding}}, \bibinfo {author} {\bibfnamefont
  {W.}~\bibnamefont {Kowalsky}}, \bibinfo {author} {\bibfnamefont
  {L.}~\bibnamefont {Kronik}}, \bibinfo {author} {\bibfnamefont {H.~J.}\
  \bibnamefont {Snaith}}, \bibinfo {author} {\bibfnamefont {A.}~\bibnamefont
  {Pucci}},\ and\ \bibinfo {author} {\bibfnamefont {R.}~\bibnamefont
  {Lovrin\v{c}i\'c}},\ }\bibfield  {title} {\bibinfo {title} {Optical phonons
  in methylammonium lead halide perovskites and implications for charge
  transport},\ }\href {https://doi.org/10.1039/C6MH00275G} {\bibfield
  {journal} {\bibinfo  {journal} {Mater. Horiz.}\ }\textbf {\bibinfo {volume}
  {3}},\ \bibinfo {pages} {613} (\bibinfo {year} {2016})}\BibitemShut {NoStop}%
\bibitem [{\citenamefont {Wright}\ \emph {et~al.}(2016)\citenamefont {Wright},
  \citenamefont {Verdi}, \citenamefont {Milot}, \citenamefont {Eperon},
  \citenamefont {P\'erez-Osorio}, \citenamefont {Snaith}, \citenamefont
  {Giustino}, \citenamefont {Johnston},\ and\ \citenamefont
  {Herz}}]{WVMEPSGJH2016}%
  \BibitemOpen
  \bibfield  {author} {\bibinfo {author} {\bibfnamefont {A.~D.}\ \bibnamefont
  {Wright}}, \bibinfo {author} {\bibfnamefont {C.}~\bibnamefont {Verdi}},
  \bibinfo {author} {\bibfnamefont {R.~L.}\ \bibnamefont {Milot}}, \bibinfo
  {author} {\bibfnamefont {G.~E.}\ \bibnamefont {Eperon}}, \bibinfo {author}
  {\bibfnamefont {M.~A.}\ \bibnamefont {P\'erez-Osorio}}, \bibinfo {author}
  {\bibfnamefont {H.~J.}\ \bibnamefont {Snaith}}, \bibinfo {author}
  {\bibfnamefont {F.}~\bibnamefont {Giustino}}, \bibinfo {author}
  {\bibfnamefont {M.~B.}\ \bibnamefont {Johnston}},\ and\ \bibinfo {author}
  {\bibfnamefont {L.~M.}\ \bibnamefont {Herz}},\ }\bibfield  {title} {\bibinfo
  {title} {Electron–phonon coupling in hybrid lead halide perovskites},\
  }\href {https://doi.org/10.1038/ncomms11755} {\bibfield  {journal} {\bibinfo
  {journal} {Nature Communications}\ }\textbf {\bibinfo {volume} {7}},\
  \bibinfo {pages} {11755} (\bibinfo {year} {2016})}\BibitemShut {NoStop}%
\bibitem [{\citenamefont {Yamada}\ and\ \citenamefont
  {Kanemitsu}(2022)}]{Yamada2022}%
  \BibitemOpen
  \bibfield  {author} {\bibinfo {author} {\bibfnamefont {Y.}~\bibnamefont
  {Yamada}}\ and\ \bibinfo {author} {\bibfnamefont {Y.}~\bibnamefont
  {Kanemitsu}},\ }\bibfield  {title} {\bibinfo {title} {Electron-phonon
  interactions in halide perovskites},\ }\href
  {https://doi.org/10.1038/s41427-022-00394-4} {\bibfield  {journal} {\bibinfo
  {journal} {NPG Asia Materials}\ }\textbf {\bibinfo {volume} {14}},\ \bibinfo
  {pages} {48} (\bibinfo {year} {2022})}\BibitemShut {NoStop}%
\bibitem [{\citenamefont {Seyitliyev}\ \emph {et~al.}(2023)\citenamefont
  {Seyitliyev}, \citenamefont {Qin}, \citenamefont {Jana}, \citenamefont
  {Janke}, \citenamefont {Zhong}, \citenamefont {You}, \citenamefont {Mitzi},
  \citenamefont {Blum},\ and\ \citenamefont
  {Gundogdu}}]{Gundogdu2023Adv_Func_Mat}%
  \BibitemOpen
  \bibfield  {author} {\bibinfo {author} {\bibfnamefont {D.}~\bibnamefont
  {Seyitliyev}}, \bibinfo {author} {\bibfnamefont {X.}~\bibnamefont {Qin}},
  \bibinfo {author} {\bibfnamefont {M.~K.}\ \bibnamefont {Jana}}, \bibinfo
  {author} {\bibfnamefont {S.~M.}\ \bibnamefont {Janke}}, \bibinfo {author}
  {\bibfnamefont {X.}~\bibnamefont {Zhong}}, \bibinfo {author} {\bibfnamefont
  {W.}~\bibnamefont {You}}, \bibinfo {author} {\bibfnamefont {D.~B.}\
  \bibnamefont {Mitzi}}, \bibinfo {author} {\bibfnamefont {V.}~\bibnamefont
  {Blum}},\ and\ \bibinfo {author} {\bibfnamefont {K.}~\bibnamefont
  {Gundogdu}},\ }\bibfield  {title} {\bibinfo {title} {Coherent phonon-induced
  modulation of charge transfer in 2d hybrid perovskites},\ }\href
  {https://doi.org/https://doi.org/10.1002/adfm.202213021} {\bibfield
  {journal} {\bibinfo  {journal} {Advanced Functional Materials}\ }\textbf
  {\bibinfo {volume} {33}},\ \bibinfo {pages} {2213021} (\bibinfo {year}
  {2023})}\BibitemShut {NoStop}%
\bibitem [{\citenamefont {Liz\'arraga}\ \emph {et~al.}(2025)\citenamefont
  {Liz\'arraga}, \citenamefont {Guerra}, \citenamefont {Enrique-Moran},
  \citenamefont {Serquen}, \citenamefont {Ventura}, \citenamefont {Villegas},
  \citenamefont {Rocha},\ and\ \citenamefont {Venezuela}}]{3sxd-26yj}%
  \BibitemOpen
  \bibfield  {author} {\bibinfo {author} {\bibfnamefont {K.}~\bibnamefont
  {Liz\'arraga}}, \bibinfo {author} {\bibfnamefont {J.~A.}\ \bibnamefont
  {Guerra}}, \bibinfo {author} {\bibfnamefont {L.~A.}\ \bibnamefont
  {Enrique-Moran}}, \bibinfo {author} {\bibfnamefont {E.}~\bibnamefont
  {Serquen}}, \bibinfo {author} {\bibfnamefont {E.}~\bibnamefont {Ventura}},
  \bibinfo {author} {\bibfnamefont {C.~E.~P.}\ \bibnamefont {Villegas}},
  \bibinfo {author} {\bibfnamefont {A.~R.}\ \bibnamefont {Rocha}},\ and\
  \bibinfo {author} {\bibfnamefont {P.}~\bibnamefont {Venezuela}},\ }\bibfield
  {title} {\bibinfo {title} {Determining exciton binding energy and reduced
  effective mass in metal tri-halide perovskites from optical and impedance
  spectroscopy measurements},\ }\href {https://doi.org/10.1103/3sxd-26yj}
  {\bibfield  {journal} {\bibinfo  {journal} {Phys. Rev. Mater.}\ }\textbf
  {\bibinfo {volume} {9}},\ \bibinfo {pages} {103806} (\bibinfo {year}
  {2025})}\BibitemShut {NoStop}%
\bibitem [{\citenamefont {Adachi}(1985)}]{GaAs1985}%
  \BibitemOpen
  \bibfield  {author} {\bibinfo {author} {\bibfnamefont {S.}~\bibnamefont
  {Adachi}},\ }\bibfield  {title} {\bibinfo {title} {Gaas, alas, and
  al${}_x$ga${}_{1-x}$as: Material parameters for use in research and device
  applications},\ }\href {https://doi.org/10.1063/1.336070} {\bibfield
  {journal} {\bibinfo  {journal} {Journal of Applied Physics}\ }\textbf
  {\bibinfo {volume} {58}},\ \bibinfo {pages} {R1} (\bibinfo {year}
  {1985})}\BibitemShut {NoStop}%
\bibitem [{\citenamefont {Leonardi}\ \emph {et~al.}(1986)\citenamefont
  {Leonardi}, \citenamefont {Persico},\ and\ \citenamefont
  {Vetri}}]{Leonardi1986DickeMA}%
  \BibitemOpen
  \bibfield  {author} {\bibinfo {author} {\bibfnamefont {C.}~\bibnamefont
  {Leonardi}}, \bibinfo {author} {\bibfnamefont {F.}~\bibnamefont {Persico}},\
  and\ \bibinfo {author} {\bibfnamefont {G.}~\bibnamefont {Vetri}},\ }\bibfield
   {title} {\bibinfo {title} {Dicke model and the theory of driven and
  spontaneous emission},\ }\href
  {https://api.semanticscholar.org/CorpusID:120444837} {\bibfield  {journal}
  {\bibinfo  {journal} {La Rivista del Nuovo Cimento (1978-1999)}\ }\textbf
  {\bibinfo {volume} {9}},\ \bibinfo {pages} {1} (\bibinfo {year}
  {1986})}\BibitemShut {NoStop}%
\bibitem [{\citenamefont {Andreev}\ \emph {et~al.}(1980)\citenamefont
  {Andreev}, \citenamefont {Emel’yanov},\ and\ \citenamefont
  {Il’inskii}}]{Andreev1980CollectiveSE}%
  \BibitemOpen
  \bibfield  {author} {\bibinfo {author} {\bibfnamefont {A.~V.}\ \bibnamefont
  {Andreev}}, \bibinfo {author} {\bibfnamefont {V.~I.}\ \bibnamefont
  {Emel’yanov}},\ and\ \bibinfo {author} {\bibfnamefont {Y.~A.}\ \bibnamefont
  {Il’inskii}},\ }\bibfield  {title} {\bibinfo {title} {Collective
  spontaneous emission (dicke superradiance)},\ }\href
  {https://api.semanticscholar.org/CorpusID:122020443} {\bibfield  {journal}
  {\bibinfo  {journal} {Physics-Uspekhi}\ }\textbf {\bibinfo {volume} {23}},\
  \bibinfo {pages} {493} (\bibinfo {year} {1980})}\BibitemShut {NoStop}%
\bibitem [{\citenamefont {Protsenko}\ and\ \citenamefont
  {Uskov}(2015)}]{Protsenko_2015}%
  \BibitemOpen
  \bibfield  {author} {\bibinfo {author} {\bibfnamefont {I.}~\bibnamefont
  {Protsenko}}\ and\ \bibinfo {author} {\bibfnamefont {A.}~\bibnamefont
  {Uskov}},\ }\bibfield  {title} {\bibinfo {title} {Superradiance of several
  atoms near a metal nanosphere},\ }\href
  {https://doi.org/10.1070/QE2015v045n06ABEH015675} {\bibfield  {journal}
  {\bibinfo  {journal} {Quantum Electronics}\ }\textbf {\bibinfo {volume}
  {45}},\ \bibinfo {pages} {561} (\bibinfo {year} {2015})}\BibitemShut
  {NoStop}%
\bibitem [{\citenamefont {Meier}\ \emph {et~al.}(1997)\citenamefont {Meier},
  \citenamefont {Zhao}, \citenamefont {Chernyak},\ and\ \citenamefont
  {Mukamel}}]{Mukamel3961997}%
  \BibitemOpen
  \bibfield  {author} {\bibinfo {author} {\bibfnamefont {T.}~\bibnamefont
  {Meier}}, \bibinfo {author} {\bibfnamefont {Y.}~\bibnamefont {Zhao}},
  \bibinfo {author} {\bibfnamefont {V.}~\bibnamefont {Chernyak}},\ and\
  \bibinfo {author} {\bibfnamefont {S.}~\bibnamefont {Mukamel}},\ }\bibfield
  {title} {\bibinfo {title} {Polarons, localization, and excitonic coherence in
  superradiance of biological antenna complexes},\ }\href
  {https://doi.org/10.1063/1.474746} {\bibfield  {journal} {\bibinfo  {journal}
  {Journal of Chemical Physics}\ }\textbf {\bibinfo {volume} {107}},\ \bibinfo
  {pages} {3876} (\bibinfo {year} {1997})}\BibitemShut {NoStop}%
\bibitem [{\citenamefont {Scully}\ and\ \citenamefont
  {Zubairy}(1997)}]{scully1997quantum}%
  \BibitemOpen
  \bibfield  {author} {\bibinfo {author} {\bibfnamefont {M.}~\bibnamefont
  {Scully}}\ and\ \bibinfo {author} {\bibfnamefont {M.}~\bibnamefont
  {Zubairy}},\ }\href@noop {} {\emph {\bibinfo {title} {Quantum Optics}}},\
  Quantum Optics\ (\bibinfo  {publisher} {Cambridge University Press},\
  \bibinfo {year} {1997})\BibitemShut {NoStop}%
\bibitem [{\citenamefont {Biliroglu}\ \emph {et~al.}(2025)\citenamefont
  {Biliroglu}, \citenamefont {T\"ure}, \citenamefont {Ghita}, \citenamefont
  {Kotyrov}, \citenamefont {Qin}, \citenamefont {Seyitliyev}, \citenamefont
  {Phonthiptokun}, \citenamefont {Abdelsamei}, \citenamefont {Chai},
  \citenamefont {Su}, \citenamefont {Herath}, \citenamefont {Swan},
  \citenamefont {Temnov}, \citenamefont {Blum}, \citenamefont {So},\ and\
  \citenamefont {Gundogdu}}]{GSoliton}%
  \BibitemOpen
  \bibfield  {author} {\bibinfo {author} {\bibfnamefont {M.}~\bibnamefont
  {Biliroglu}}, \bibinfo {author} {\bibfnamefont {M.}~\bibnamefont {T\"ure}},
  \bibinfo {author} {\bibfnamefont {A.}~\bibnamefont {Ghita}}, \bibinfo
  {author} {\bibfnamefont {M.}~\bibnamefont {Kotyrov}}, \bibinfo {author}
  {\bibfnamefont {X.}~\bibnamefont {Qin}}, \bibinfo {author} {\bibfnamefont
  {D.}~\bibnamefont {Seyitliyev}}, \bibinfo {author} {\bibfnamefont
  {N.}~\bibnamefont {Phonthiptokun}}, \bibinfo {author} {\bibfnamefont
  {M.}~\bibnamefont {Abdelsamei}}, \bibinfo {author} {\bibfnamefont
  {J.}~\bibnamefont {Chai}}, \bibinfo {author} {\bibfnamefont {R.}~\bibnamefont
  {Su}}, \bibinfo {author} {\bibfnamefont {U.}~\bibnamefont {Herath}}, \bibinfo
  {author} {\bibfnamefont {A.}~\bibnamefont {Swan}}, \bibinfo {author}
  {\bibfnamefont {V.~V.}\ \bibnamefont {Temnov}}, \bibinfo {author}
  {\bibfnamefont {V.}~\bibnamefont {Blum}}, \bibinfo {author} {\bibfnamefont
  {F.}~\bibnamefont {So}},\ and\ \bibinfo {author} {\bibfnamefont
  {K.}~\bibnamefont {Gundogdu}},\ }\bibfield  {title} {\bibinfo {title}
  {Unconventional solitonic high-temperature superfluorescence from
  perovskites},\ }\href {https://doi.org/10.1038/s41586-025-09030-x} {\bibfield
   {journal} {\bibinfo  {journal} {Nature}\ }\textbf {\bibinfo {volume}
  {642}},\ \bibinfo {pages} {71} (\bibinfo {year} {2025})}\BibitemShut
  {NoStop}%
\bibitem [{\citenamefont {Gladkij}\ \emph {et~al.}(2025)\citenamefont
  {Gladkij}, \citenamefont {Veretenov}, \citenamefont {Rosanov}, \citenamefont
  {Malomed}, \citenamefont {Osipov},\ and\ \citenamefont
  {Fainberg}}]{gladkij2025stabilityquantumcoherentsuperradiant}%
  \BibitemOpen
  \bibfield  {author} {\bibinfo {author} {\bibfnamefont {A.~A.}\ \bibnamefont
  {Gladkij}}, \bibinfo {author} {\bibfnamefont {N.~A.}\ \bibnamefont
  {Veretenov}}, \bibinfo {author} {\bibfnamefont {N.~N.}\ \bibnamefont
  {Rosanov}}, \bibinfo {author} {\bibfnamefont {B.~A.}\ \bibnamefont
  {Malomed}}, \bibinfo {author} {\bibfnamefont {V.~A.}\ \bibnamefont
  {Osipov}},\ and\ \bibinfo {author} {\bibfnamefont {B.~D.}\ \bibnamefont
  {Fainberg}},\ }\href {https://arxiv.org/abs/2511.03600} {\bibinfo {title}
  {Stability of the quantum coherent superradiant states in relation to
  exciton-phonon interactions and the fundamental soliton in hybrid
  perovskites}} (\bibinfo {year} {2025}),\ \Eprint
  {https://arxiv.org/abs/2511.03600} {arXiv:2511.03600 [nlin.PS]} \BibitemShut
  {NoStop}%
\end{thebibliography}
\end{document}